\documentclass[%
 reprint,
 superscriptaddress,
 preprintnumbers,
 nofootinbib,
 amsmath,amssymb,
 aps,
 prl,
]{revtex4-2}

\usepackage{algorithm}
\usepackage{algpseudocode}
\usepackage{psfrag,slashed,cancel,array,graphicx,todonotes}
\usepackage{subfigure}
\usepackage[labelfont=bf,labelsep=quad,justification=justified]{caption}
\usepackage[utf8]{inputenc}
\usepackage{mathtools}
\usepackage{mathrsfs}
\usepackage{multirow}
\usepackage{braket}
\usepackage{slashed}
\usepackage{booktabs} 
\usepackage[normalem]{ulem}
\usepackage{slashed}
\usepackage{hyperref}

\hypersetup{pdftitle={The perturbative structure of vetoes in azimuthal gaps},pdfcreator={},linkcolor=[rgb]{0.15,0.35,0.75},colorlinks=true,citecolor=[rgb]{0.675,0,0.2},urlcolor=[rgb]{0.15,0.35,0.65}}

\newcommand{\as}{\alpha_s}

\newcommand{\X}{{\boldsymbol{X}}}
\newcommand{\T}{{\boldsymbol{T}}}
\newcommand{\G}{{\boldsymbol{\Gamma}}}
\newcommand{\D}{{\boldsymbol{D}}}

\newcommand{\ymax}[1]{{\eta_{#1}^{\mathrm{max}}}}
\newcommand{\com}[2]{{\left[{#1},{#2}\right]}}

\newcommand{\asfact}{{\frac{2\alpha_s}{\pi}}}

\def\beq{\begin{equation}}
\def\eeq{\end{equation}}
\def\bea{\begin{eqnarray}}
\def\eea{\end{eqnarray}}

\AtBeginDocument{
\heavyrulewidth=.08em
\lightrulewidth=.05em
\cmidrulewidth=.03em
\belowrulesep=.65ex
\belowbottomsep=0pt
\aboverulesep=.4ex
\abovetopsep=0pt
\cmidrulesep=\doublerulesep
\cmidrulekern=.5em
\defaultaddspace=.5em
}

\newcommand{\EQ}{Eq.~}

\def\be{\begin{equation}}
\def\ee{\end{equation}}

\begin{document}

\preprint{CERN-TH-2025-209}

\title{Uncharted Logarithmic Structures in QCD Transverse-Energy Flow}

\author{Mrinal Dasgupta}
\email{mrinal.dasgupta@manchester.ac.uk}
\affiliation{Department of Physics \& Astronomy, University of Manchester, Manchester M13 9PL, United Kingdom}

\author{Alexander Fraley}
\email{alexander.fraley@postgrad.manchester.ac.uk}
\affiliation{Department of Physics \& Astronomy, University of Manchester, Manchester M13 9PL, United Kingdom}

\author{Pier Francesco Monni}
\email{pier.monni@cern.ch}
\affiliation{CERN, Theoretical Physics Department, CH-1211 Geneva 23, Switzerland}

\author{Saad Nabeebaccus}
\email{saad.nabeebaccus@manchester.ac.uk}
\affiliation{Department of Physics \& Astronomy, University of Manchester, Manchester M13 9PL, United Kingdom}

\begin{abstract}
We investigate the QCD transverse-energy ($E_T$) flow distribution within an azimuthal region of phase space, defined by an angular interval $\Delta \phi$ on the plane transverse to a chosen jet axis. Vetoes on the resulting $E_T$ are widely employed at the LHC to isolate missing transverse momentum in final states with invisible particles. We show that this observable has logarithmic structures never seen before in QCD calculations. Notably, its description involves the resummation of non-global and coherence-violating logarithmic contributions that are more singular than any reported to date. We analyze its all-orders behavior, providing analytical resummations at next-to-double-logarithmic accuracy for $e^+e^-$ and $pp$ collisions, and numerical resummations at leading-logarithmic accuracy in the Veneziano limit for $pp$ collisions. We further present a computation of the leading coherence-violating correction for $pp$ collisions. The intricate structure of vetoes in azimuthal gaps reveals new aspects of QCD dynamics and offers a novel probe of collinear-factorization breaking at the LHC.
\end{abstract}

\maketitle

\paragraph*{Introduction.---} In the past decades, the outstanding physics program of the Large Hadron Collider (LHC) has triggered a new wave of theoretical progress aimed at describing the complex structure of scattering observables with an unprecedented level of precision. These advancements concern different areas of quantum field theory, notably Quantum Chromodynamics (QCD), and are instrumental in deepening our understanding of non-Abelian gauge theories as well as optimizing the physics potential of the LHC and future collider experiments.
In spite of the remarkable progress of the past years, the description of the rich structure of QCD dynamics in real-life scattering observables still constitutes, in general, a formidable problem. The use of exclusive kinematic cuts in experimental measurements often challenges the ability to make precise predictions for multi-scale observables. An important example is that of non-global observables~\cite{Dasgupta:2001sh,Banfi:2002hw}. These measure the QCD activity in selected regions of the solid angle, designed in order to enhance the sensitivity to a specific physics effect of interest. 
Arguably the best-known variant of this class of observable~\cite{Dasgupta:2002bw} is the measurement, in $e^+e^-\to 2$ jets, of the fraction of events that survive a veto $E_T$ on the energy flow transverse to a reference axis $\hat{n}$ (typically the thrust axis) measured inside a region $\Omega$ of the solid angle
\begin{equation}\label{eq:master}
    \Sigma = \frac{1}{\sigma_0}\int_0^{E_T}\!\!d E_{T,\Omega}\frac{d\sigma}{d E_{T,\Omega}},\quad E_{T,\Omega} = \sum_{k_i\in \Omega} k_{t, i},
\end{equation}
where $\sigma_0$ is the Born cross section for the process under consideration, and the $k_{t, i}$'s are the transverse momenta with respect to $\hat{n}$.
In the  limit where the veto scale $E_T$ is much smaller than the COM scale $Q$ of the underlying hard scattering process, the perturbative expansion of $\Sigma$ features large radiative corrections enhanced by logarithms $L\equiv \ln Q/E_T \gg 1$. The presence of these large corrections impairs the convergence of the perturbative prediction and demands the all-order resummation of the corresponding series. The resummed perturbative expansion can be organized in two ways, at the level of~$\Sigma$
\begin{equation}\label{eq:sigma_counting}
    \Sigma = 1 + g_{\scriptscriptstyle{\rm DL}}(\alpha_s L^2) + \frac{1}{L} g_{\scriptscriptstyle{\rm NDL}}(\alpha_s L^2)+ \dots,
\end{equation}
or, for observables that satisfy the property of \textit{exponentiation}, at the level of $\ln\Sigma$
\begin{equation}\label{eq:ln_sigma_counting}
\ln\Sigma = L g_{\scriptscriptstyle{\rm LL}}(\alpha_s L) + g_{\scriptscriptstyle{\rm NLL}}(\alpha_s L)+ \frac{1}{L} g_{\scriptscriptstyle{\rm NNLL}}(\alpha_s L)+ \dots\,.
\end{equation}
The former counting is equivalent to expanding $\Sigma$ in powers of $\sqrt{\alpha_s}$ at fixed $\alpha_s L^2\sim 1$, and the latter to expanding $\ln\Sigma$ in $\alpha_s$ at fixed $\alpha_s L\sim 1$.
In the example of Eq.~\eqref{eq:master}, the region $\Omega$ is typically defined by an angular cut away from the axis $\hat{n}$ along which the large energy flow of the event propagates. This is achieved by requiring that the rapidity of particles $|\eta| < \eta_{\rm cut}\sim {\cal O}(1)$ and, optionally, by an additional cut on the azimuthal angle $\phi$. 
In this case the large logarithmic corrections originate from soft partons propagating with a wide angle with respect to $\hat{n}$, and for this reason the series of $\Sigma$ is said to be \textit{single-logarithmic}, i.e.~double-logarithmic (DL) terms are absent in Eqs.~\eqref{eq:sigma_counting},~\eqref{eq:ln_sigma_counting} ($g_{\scriptscriptstyle{\rm DL}}(\alpha_s L^2)=g_{\scriptscriptstyle{\rm LL}}(\alpha_s L)=0$).

For such observables, state of the art resummations, commonly obtained using numerical methods based on the large-$N_c$ limit of QCD, reach next-to-next-to-leading-logarithmic (NNLL) accuracy~\cite{Caron-Huot:2015bja,Banfi:2021owj,Banfi:2021owj,Becher:2021urs,Becher:2023vrh,FerrarioRavasio:2023kyg}, i.e.~they control terms up to and including the function $g_{\scriptscriptstyle{\rm NNLL}}(\alpha_s L)$ in the $\ln\Sigma$ counting of Eq.~\eqref{eq:ln_sigma_counting}. Furthermore, in the NLL approximation, a variety of calculations were performed in the literature, both at high fixed perturbative orders~\cite{Schwartz:2014wha,Khelifa-Kerfa:2015mma,Khelifa-Kerfa:2024udm} as well as at all orders in full color~\cite{Hatta:2013iba,Hagiwara:2015bia,Nagy:2019bsj,Hatta:2020wre,DeAngelis:2020rvq}. This progress in understanding non-global observables resulted in several key applications to collider phenomenology (cf., e.g.,~\cite{Rubin:2010fc,Banfi:2010pa,Dasgupta:2012hg,Kerfa:2012yae,Becher:2015hka,Becher:2016omr,Neill:2016stq,Larkoski:2015zka,Larkoski:2016zzc,Becher:2017nof,AngelesMartinez:2018cfz,Balsiger:2018ezi,Neill:2018yet,Balsiger:2019tne,Balsiger:2020ogy,Ziani:2021dxr,Larkoski:2025afg}).

In this Letter, we initiate a study of the above observable in a different regime, that is when $L \ll \eta_{\rm cut}$, effectively removing any rapidity constraint on emissions, while retaining a cut on the azimuthal angle w.r.t.~the reference axis $\hat{n}$, i.e.~$|\phi| < \Delta\phi/2$. As we will show, this seemingly simple modification of the observable introduces considerable complexity in the perturbative resummation that challenges the state-of-the-art methods used to tackle these calculations. One example of these challenges is an enhanced sensitivity to non-global logarithms, showing up already at the $g_{\scriptscriptstyle{\rm NDL}}(\alpha_s L^2)$ and $g_{\scriptscriptstyle{\rm LL}}(\alpha_s L)$ level in Eqs.~\eqref{eq:sigma_counting},~\eqref{eq:ln_sigma_counting}, respectively. 

Our investigation has important phenomenological implications. Indeed, such vetoes in azimuthal gaps are often used as an isolation cut in measurements involving missing transverse momentum at the LHC (for instance in mono-jet Dark Matter searches~\cite{CMS:2021far,ATLAS:2021kxv,ATLAS:2024vqf}). The precise understanding of this type of cut is therefore paramount to robustly predict the structure of the Standard Model in these key applications.
Furthermore, at hadron colliders the increased sensitivity to non-global logarithms necessarily implies the presence of large factorization-violating effects, making the observable introduced here a suitable probe for their study.

\paragraph*{The lepton-collider case.---}
We start by analyzing the lepton-collider case, where a $q\bar{q}$ pair is produced in an $e^+e^-$ collision of COM energy $Q$. It is instructive to examine first the structure of the observable in the most singular configurations, namely when the final-state $q\bar{q}$ pair is accompanied by soft-collinear primary emissions that are strongly ordered both in energy and relative angle. This picture is accurate at the double-logarithmic (DL) order ($g_{\scriptscriptstyle{\rm DL}}(\alpha_s L^2)$). In this limit, most known observables exhibit an exponential structure that fixes all terms in $g_{\scriptscriptstyle{\rm DL}}(\alpha_s L^2)$ in terms of the leading-order, ${\cal O}(\alpha_s)$ one. Here, sensitivity to the reference axis $\hat{n}$ makes the situation more subtle. To see this, let us take $\hat{n}$ to be the widely used thrust axis~\cite{Brandt:1964sa,Farhi:1977sg}. Crucially, the only phase-space constraint at first order is on the azimuth of the first emission relative to $\hat{n}$. In particular, there is a veto on its transverse momentum when it enters, or recoils the quark into, the azimuthal gap $\Omega$. The existence of a recoil contribution from the quark doubles the effective size of $\Omega$ and so we consider only $\Delta \phi<\pi$ for the thrust axis variant of the observable. The rapidity of the emission is integrated over inclusively up to its kinematic limit $|\eta| \leq \ln Q/k_t$. To leading order, the DL cross-section reads
$\Sigma^{\scriptscriptstyle{\rm (DL)}}_{\scriptscriptstyle{\rm Thrust}} = 1 - 8 C_F f \bar{\alpha} L^2  + \dots$, with $f\equiv \Delta\phi/2\pi$ and $\bar{\alpha}\equiv \alpha_s/2\pi$.

\begin{figure}[h]
     \centering
     \includegraphics[width=0.6\linewidth]{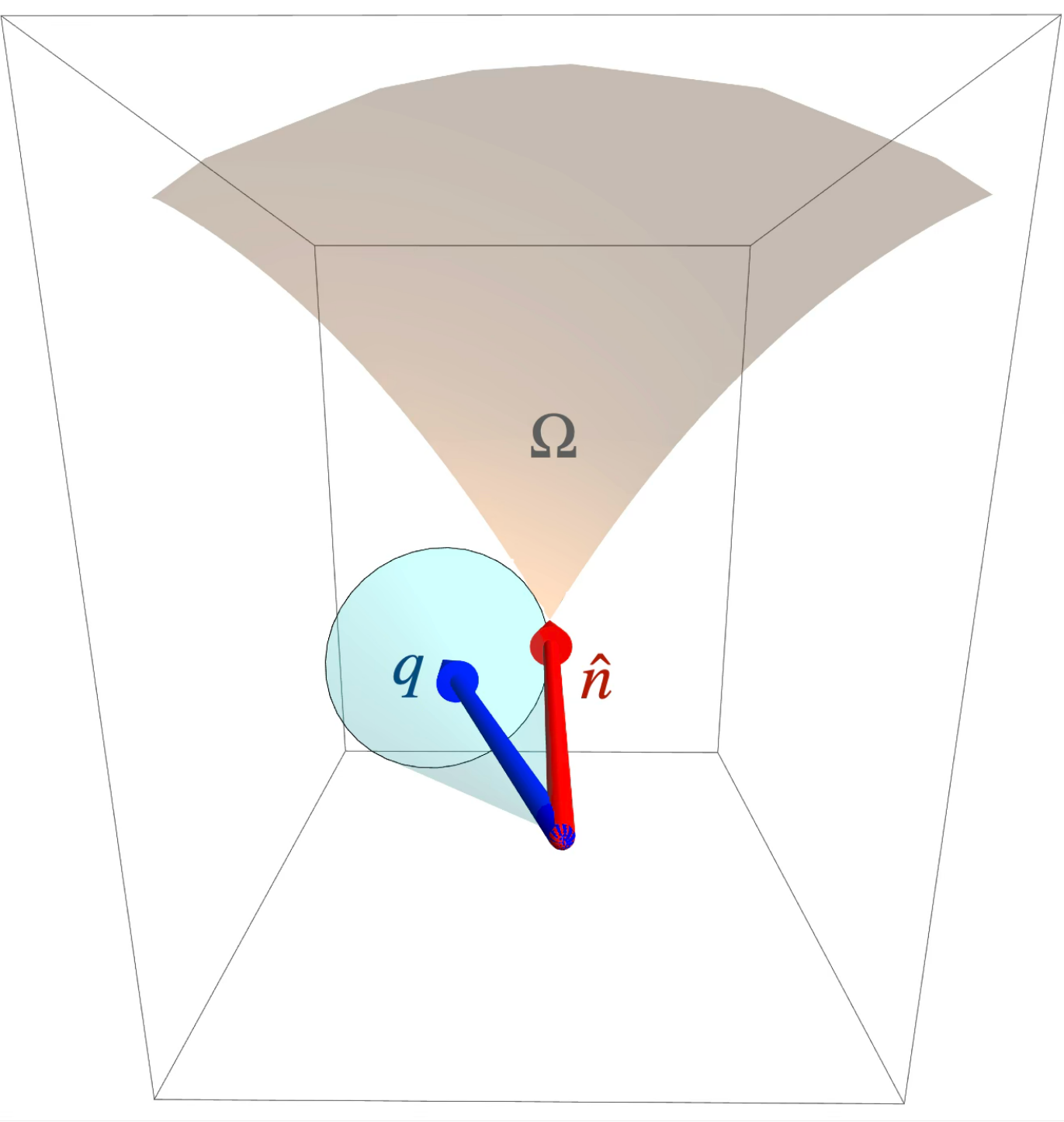}
     \caption{Configuration in which the hard quark (blue arrow) is misaligned with the reference axis $\hat{n}$ (red arrow).}
     \label{fig:tilt}
 \end{figure}

At second order, we consider the emission of two soft-collinear gluons $k_1$, $k_2$ (with $k_{t,1}\gg k_{t,2}$) from the final state quark (or anti-quark). The emission of $k_1$ causes the hard quark to recoil and as a result the quark will not be aligned with the thrust axis any longer, as displayed in Fig.~\ref{fig:tilt}. 
At this point the azimuth-only constraint for a single emission does not carry over to $k_2$. Although $k_2$ can reach $\eta_2=\ln Q/k_{t,2}$ kinematically, it requires a veto only when $\eta_2<-\ln \theta_{\hat{n} q}$, with $\theta_{\hat{n} q}\approx k_{t,1}/Q$ being the quark-$\hat{n}$ angle. The region of rapidity that no longer requires a veto at DL accuracy is displayed by the blue cone in Fig.~\ref{fig:tilt}. The misalignment between the emitting quark and axis $\hat{n}$ is a persistent effect in this Letter and we refer to it generally as \textit{tilt}. This phase space entanglement between two strongly-ordered emissions breaks the simple exponentiation pattern already at DL accuracy. At next-to--leading order we find that $\Sigma^{\scriptscriptstyle{\rm (DL)}}_{\scriptscriptstyle{\rm Thrust}}$ reads
\begin{equation}
\Sigma^{\scriptscriptstyle{\rm (DL)}}_{\scriptscriptstyle{\rm Thrust}} = 1 - 8 C_F f \bar{\alpha} L^2  + \frac{16}{3}C_F^2 f(1+4f)\bar{\alpha}^2 L^4+ \dots,    
\end{equation}
clearly demonstrating DL exponentiation breaking, the first of a series of non-trivial features we shall discover. In spite of the lack of simple exponentiation, the resummation of DL terms does remain tractable and we report the DL resummed expression in Ref.~\cite{supplemental},~\S\ref{app:DL-thrust}. 

Going beyond DL with this choice of axis entails handling several complications and hence we do not pursue the thrust axis definition any further. We instead seek an axis definition which eliminates the problems seen at DL in the thrust case.
 A possible choice is to take $\hat{n}$ to be the WTA axis~\cite{Larkoski:2014uqa}. For configurations involving a hard quark and multiple strongly-ordered soft gluons, this axis aligns with the direction of the hardest parton, i.e.~the quark. This eliminates the recoil issues described above. The DL cross section hence arises from simple exponentiation of the leading-order result (cf. Ref.~\cite{supplemental},~\S\ref{app:DL-WTA})
 \begin{equation}
     \Sigma^{\scriptscriptstyle{\rm (DL)}}_{\scriptscriptstyle{\rm WTA}} = e^{- 4 C_F f \bar{\alpha} L^2}.
 \end{equation}
 The exponentiation of DL terms suggests that the WTA axis is a more natural definition of $\hat{n}$, and allows us to go beyond DL accuracy in the counting of Eq.~\eqref{eq:sigma_counting}. In what follows we pursue the goal of achieving NDL accuracy, in the process discovering a number of further interesting features. To this end we need to consider, in addition to an ensemble of soft-collinear emissions, the splitting of one of the emitted soft gluons as well as the limit in which an emission is hard and collinear to the quark.

 We start by analyzing the case of a gluon splitting into either two soft gluons or a pair of soft quarks. If the two daughters are inside $\Omega$, this configuration gives rise to a standard physical-coupling scheme~\cite{Catani:1990rr,Banfi:2018mcq,Catani:2019rvy}. On the other hand, a second possibility is that the harder of the two daughters ends up outside $\Omega$ while the softer ends up inside. 
 Such configurations are also well known, giving rise to the classic non-global logarithms, which in general are beyond NDL effects. However in the present case a detailed analysis gives rise to a surprising result: a non-global logarithmic correction appearing already at ${\cal O}(\bar{\alpha}^2 L^3)$, i.e.~at NDL. Relative to standard non-global logarithms with terms starting from ${\cal O}(\bar{\alpha}^2 L^2)$, the extra $L$ appearing here can be attributed to the fact that the edge of the veto region $\Omega$ extends into the collinear limit due to the effective absence of an explicit rapidity cut. 
 
\begin{figure}[t]
     \centering
\includegraphics[width=0.9\linewidth]{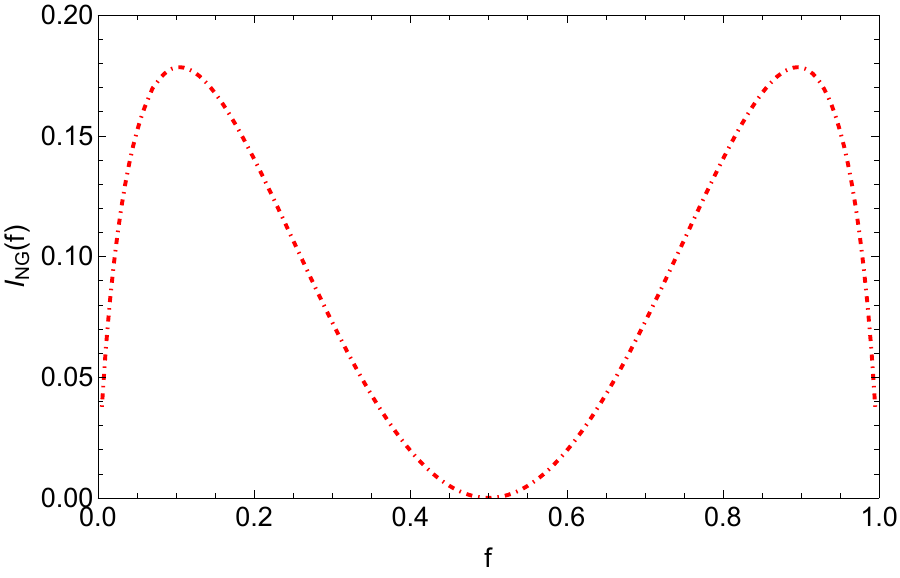}
     \caption{The quantity $I_{\scriptscriptstyle\text{NG}}(f)$ as a function of the opening of the azimuthal gap.}
     \label{fig:IngS}
 \end{figure}

 The non-global contribution needed to achieve NDL accuracy takes the form~\cite{supplemental},~\S\ref{app:NGLs-NDL}
\begin{equation}\label{eq:NLO_ng_structure}
    \Sigma^{\scriptscriptstyle{\rm (NDL)}}_{\scriptscriptstyle\text{NG}} =-\frac{8}{3}C_FC_AI_{\scriptscriptstyle\text{NG}}(f)\Bar{\alpha}^2L^3,
\end{equation} 
where
\begin{equation}\label{eq:explicit_IngS}
    I_{\scriptscriptstyle\text{NG}}(f)=\!\!\int_{k_{1}\notin\Omega}\!\!\frac{d\phi_1}{2\pi}\int_{k_{2}\in\Omega}\!\!\frac{d\phi_2}{2\pi}\!\!\int_{-\infty}^{\infty}\!\!d\eta \frac{{\rm c}(\phi_{12})}{{\rm ch}(\eta)-{\rm c}(\phi_{12})},
\end{equation}
and ${\rm ch}(\eta)\equiv \cosh(\eta)= \cosh(\eta_1-\eta_2)$ and ${\rm c}(\phi_{12})\equiv\cos( \phi_1-\phi_2)$. The result is shown in Fig.~\ref{fig:IngS} as a function of $f$. We observe that the non-global correction vanishes for $f=1/2$ ($\Delta\phi=\pi$) at NDL.
 The appearance of non-global logarithms already at NDL level, a remarkable novelty of the present observable, implies a greater sensitivity to soft dynamics which complicates significantly its resummation structure beyond NDL.

To achieve full NDL accuracy the only other configuration that we must consider is the emission of a hard-collinear gluon from the quarks, accompanied by any number of soft-collinear gluons. This gives rise to two types of contribution to the observable. The first is when the hard collinear gluon is softer than the emitting quark, which implies that the WTA axis is still aligned with the quark. This contribution simply requires account of the hard-collinear term in the $q\to qg$ splitting, encoded in the standard $P_{gq}(z)$ splitting function~\cite{Gribov:1972ri,Altarelli:1977zs,Dokshitzer:1977sg}.

The second, more subtle case, is when the gluon is the harder parton, and it causes the WTA axis $\hat{n}$ to move along its direction rather than that of the quark. The region $\Omega$ is now defined w.r.t.~the gluon direction and we must veto emissions from both the quark and secondary emissions from the hard gluon. For emissions off the quark, the situation is analogous to that discussed for the thrust-axis case since $\hat{n}$ is now misaligned with the quark, as illustrated in Fig.~\ref{fig:tilt}.
Unlike for the thrust-axis case, for the WTA case the \textit{tilt} of the reference axis is caused by a hard-collinear emission, hence postponing the effect to NDL order. 

Putting the above two cases together results in the following NDL resummation for $\Sigma$ in the WTA-axis case (full details are given in Ref.~\cite{supplemental},~\S\ref{app:DL-WTA}), 
\begin{widetext}
\begin{multline}\label{Eq:WTA resummation result}
    \Sigma^{\scriptscriptstyle{\rm (NDL)}}_{\scriptscriptstyle{\rm WTA}}=e^{-R(L)}
    \bigg(1
    -\tfrac{8}{3}C_FC_AI_{\scriptscriptstyle\text{NG}}(f)\bar{\alpha}^2L^3\bigg)
    +\, \frac{1}{2}\,C_F\,\bar{\alpha}\,e^{-R(L)}\,
    (1-f)\,(16\ln 2-5) \left(\sqrt{\pi}\frac{\mathrm{erf}(\xi L)}
    {2\xi}-L\right)\,
    ,
\end{multline}
\end{widetext}
where $R(L)$ is the standard Sudakov radiator
\begin{equation}
    R(L) = 2 C_F f \int^{L}_{0} d \ell \frac{\alpha_s(Q e^{-\ell})}{2\pi} \left(4\ell - 3\right),
\end{equation}  
and we have defined $\xi \equiv \sqrt{2\bar{\alpha} f\,(C_A-C_F)}$.
The structure of Eq.~\eqref{Eq:WTA resummation result} can be understood as follows. The first term in the r.h.s. describes the naive exponentiation of the single gluon result supplemented by a non-global correction truncated at NDL accuracy. It correctly accounts for double-logarithmic terms and NDL terms of soft origin, while not completely accounting for the NDL structure from hard-collinear emissions.

The second term provides the necessary NDL correction via the tilt effect induced by a hard-collinear splitting which moves the reference axis $\hat{n}$ away from the direction of the hard quark. As described above, this term takes account of additional soft-collinear radiation from the hard quark as well as secondary emissions from the hard gluon. The presence of non-global logarithms and axis swap effects already at NDL makes resummation beyond NDL accuracy complicated in practice and hence we truncate our resummation at NDL level.

\begin{figure}[t]
     \centering
     \includegraphics[width=0.9\linewidth]{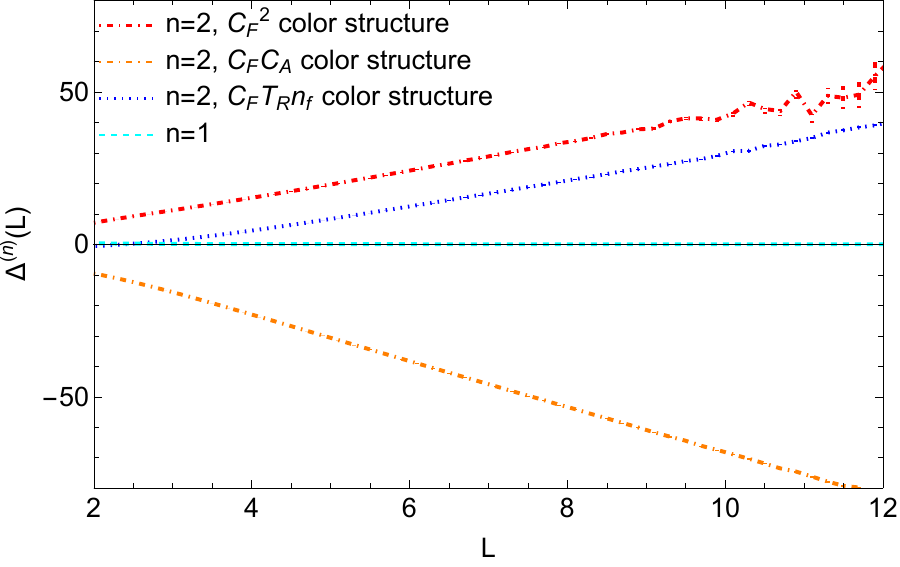}
     \caption{The quantity $\Delta^{(n)}(L)$, with $n=1,2$.}
     \label{fig:Event2}
 \end{figure}

To test the correctness of Eq.~\eqref{Eq:WTA resummation result}, we can compare its second-order, ${\cal O}(\alpha_s^2)$ expansion to a fixed-order prediction obtained with the program \texttt{Event2}~\cite{Catani:1996vz}. Fig.~\ref{fig:Event2} shows the quantity
\begin{equation}
    \Delta^{(n)}(L) \equiv \frac{d\Sigma^{(n)}_{\text{Eq.~\eqref{Eq:WTA resummation result}}}}{d L}-\frac{d\Sigma^{(n)}_{\text{\texttt Event2}}}{d L},
\end{equation}
for $f=1/4$ ($\Delta\phi=\pi/2$), with $\frac{d\Sigma}{dL} = \sum_{m=1}^2\frac{\alpha^m_s}{(2\pi)^m} \frac{d\Sigma^{(m)}}{dL}$.
At NDL, in the asymptotic limit of large $L$ one expects that $\Delta^{(1)}(L)\to 0$ and $\Delta^{(2)}(L) \propto L$. These scalings are confirmed numerically in Fig.~\ref{fig:Event2}, providing a non-trivial check of our calculation.

\paragraph*{The hadron-collider case.---} 
We now move on to study the jet-veto cross section in an azimuthal gap at hadron colliders. We consider the production of a Drell-Yan lepton pair of invariant mass $Q$ in $pp$ collisions. The observable we study is defined by vetoing QCD radiation in an azimuthal gap of width $\Delta\phi$ w.r.t.~the beam axis.

The extension to hadronic collisions entails some significant differences to the lepton-collider case.
Choosing the reference axis to be aligned with the beam axis ensures that the emitting initial-state partons are always aligned with $\hat{n}$. A direct consequence of this fact is the absence of the corrections due to the tilt effect that was responsible for significant complications in the lepton-collider variants presented above.
However, as we shall discuss, a number of other non-trivial effects emerge at hadron colliders which also impinge upon the accuracy of analytical resummation. 

The hadron-collider counterpart of the NDL result of Eq.~\eqref{Eq:WTA resummation result} simply reads
\begin{align}\label{Eq:pp resummation result}
    \Sigma^{\scriptscriptstyle{\rm (NDL)}}_{\scriptscriptstyle{\rm DY}}&=e^{-R(L)}
    \bigg(1
    -\tfrac{8}{3}C_FC_AI_{\scriptscriptstyle\text{NG}}(f)\bar{\alpha}^2L^3\bigg)\\
  & \!\!\!\!\!\!\!\!\!- 2 \bar{\alpha} C_F f e^{-R(L)} L \left(\frac{(\hat{P}^{(0)}\otimes h)_q(x_1,Q)}{h_{q}(x_1,Q)} + \{1\leftrightarrow 2\}\right),\notag
\end{align}
where $h_i(x,\mu)$ denotes the parton density for finding a parton of flavour $i$ in the proton at a scale $\mu$ and carrying a longitudinal momentum fraction $x$. $\hat{P}^{(0)}$ denotes the regularized Altarelli-Parisi leading-order splitting function~\cite{Gribov:1972ri,Altarelli:1977zs,Dokshitzer:1977sg}.
As in the lepton-collider case, the first term in Eq.~\eqref{Eq:pp resummation result} takes account of soft emissions and hard-collinear virtual terms, while the second term encodes the emission of a real initial-state, hard-collinear emission. In the latter term, the $f$ pre-factor indicates that the initial-state hard-collinear radiation is constrained by the azimuthal veto. While in a typical global observable this type of term can be reabsorbed into the scale evolution of the parton densities, the presence of the $f$ factor makes this impossible for the present observable. Going beyond NDL we encounter a number of additional issues. A first complication, that initially appears at NNDL order, is due to azimuthal correlations that in general exist between a pair of hard collinear emissions. The presence of azimuthal cuts in the $\Delta\phi$ observable definition makes us sensitive to such correlations requiring a proper treatment of spin correlations at all orders.

A second, arguably more serious complication is due to the presence of coherence-violating logarithms (CVLs) associated with violations of collinear factorization. Such logarithmic corrections are common in non-global observables at hadron-colliders, but while in known cases~\cite{Forshaw:2006fk,Forshaw:2008cq,Forshaw:2009fz,DuranDelgado:2011tp,Becher:2021zkk,Becher:2023mtx,Boer:2023jsy,Boer:2023ljq,Boer:2024hzh,Becher:2024nqc,Banerjee:2025kkq} they manifest themselves at N$^3$DL, in our case the leading contribution to the CVLs tower reads~\cite{supplemental},~\S\ref{app:CVL} 
\begin{align}
\label{eq:sigmaCVL}
    \Sigma^{\scriptscriptstyle{\rm (CVL)}}_{\scriptscriptstyle{\rm DY}} = -  \pi^2 C_F N_c^2 f\,\left(1-f\right)^2\frac{8}{45}\,\bar{\alpha}^5\,L^8,
\end{align}
that is they already contribute at NNDL, albeit coming with the usual suppression relative to leading-color terms. 
The enhancement of CVLs makes the observable introduced here valuable for investigations into collinear factorization breaking. We leave this important avenue for future work, together with the investigation of other sources of NNDL corrections.

\paragraph*{LL resummation in the hadron-collider case.---} 
The absence of tilt corrections in the hadron-collider case also enables us to use the $\ln\Sigma$ counting of Eq.~\eqref{eq:ln_sigma_counting} in the large-$N_c$ limit, where there are no complications from factorization breaking. Hence, we derive a LL prediction in the Veneziano limit, that is when $N_c\sim n_f \gg 1$.

\begin{figure}[t]
    \centering
    \includegraphics[width=0.9\linewidth]{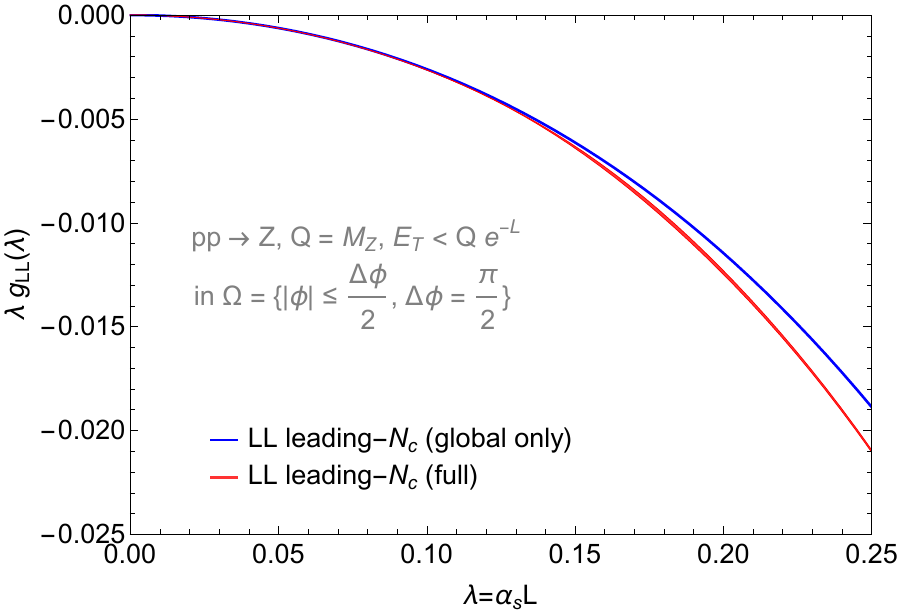}
    \caption{The function $\lambda \,g_{\scriptscriptstyle{\rm LL}}(\alpha_s L)$ vs.~the naive global result $\lambda \,g^{\scriptscriptstyle\rm global}_{\scriptscriptstyle{\rm LL}}(\alpha_s L)$. The value $\lambda\simeq 0.25$ corresponds to $E_T=30$ GeV in a scenario where $Q=500$ GeV.}
    \label{fig:gnole}
\end{figure}

Remarkably, even at LL accuracy, the $\Delta\phi$ observable requires the full resummation of the non-global tower of logarithms associated with multiple soft emissions at commensurate angles and with strong energy ordering. To achieve this, we derive a numerical resummation algorithm that we implement in the \texttt{Gnole} program~\cite{Banfi:2021owj,Banfi:2021owj}. In Ref.~\cite{supplemental},~\S\ref{app:NGL-LL} we describe in detail the core modification w.r.t.~the algorithm presented in Refs.~\cite{Dasgupta:2001sh,Banfi:2021owj}. This amounts to the implementation of the correct kinematic bound in the rapidity of the emissions, that captures the additional collinear enhancement. This step allows us to obtain a numerical prediction for the LL function $g_{\scriptscriptstyle{\rm LL}}(\alpha_s L)$ that we display in Fig.~\ref{fig:gnole}. 
The figure also shows the result due to the naive \textit{global} resummation resulting from the primary soft-collinear emissions from the quarks, which admits a simple analytical form 
\begin{equation}
g^{\scriptscriptstyle\rm global}_{\scriptscriptstyle{\rm LL}}(\lambda)  = C_F f \frac{\ln(1-2 \beta_0\lambda)+2\beta_0\lambda}{\beta_0^2 \pi \lambda},
\end{equation}
where $\lambda =\alpha_s L$, $\beta_0=(11 C_A-2 n_f)/(12\pi)$. Since we work in the Veneziano limit, we set $C_F=C_A/2$.
The difference between the two results amounts to non-global corrections, thus highlighting their sizable contribution to this observable already at LL.

Another key application of our results is in the context of developing parton showers with higher logarithmic accuracy (cf.~\cite{Dasgupta:2018nvj,Dasgupta:2020fwr,Hamilton:2020rcu,Karlberg:2021kwr,Hamilton:2021dyz,vanBeekveld:2022zhl,vanBeekveld:2022ukn,vanBeekveld:2023chs,Forshaw:2020wrq,Nagy:2020dvz,Nagy:2020rmk,Herren:2022jej,vanBeekveld:2023lsa,Assi:2023rbu,Preuss:2024vyu,Hoche:2024dee,vanBeekveld:2024qxs,vanBeekveld:2024wws,vanBeekveld:2025lpz} for recent work). In this regard, it is interesting to observe that common $k_t$-ordered dipole showers with local recoil will fail to reproduce even the LL leading-$N_c$ resummation for the $\Delta\phi$ observable. A proof of the failure of such showers at the LL level is given in the supplemental material to this Letter~\cite{supplemental},~\S\ref{app:PS}. 

\paragraph*{Outlook.---} 
In this Letter, we have initiated the exploration of non-global observables in a new kinematic regime, where the constraint $\eta_{\rm cut}$ on the rapidity of the radiation is much larger than the large logarithm $L$, such as in the case where one constrains the radiation only in an azimuthal region of phase space. This seemingly simple phase-space cut leads to profound complications for the accurate QCD resummation of this class of observables. Among the various novelties that we uncovered we highlighted the delicate dependence of the observable on the choice of axis at lepton colliders and the appearance of non-global logarithmic corrections at the LL level, as well as the enhanced contribution from CVLs at hadron colliders. Aside from its theoretical novelty, our study is relevant for experimental analyses that impose azimuthal isolation cuts for example around the missing-transverse-energy vector in new-physics searches. Here, the precise interplay between the azimuthal cut, any rapidity cuts and the veto on jets is critical for accurate theory predictions. In cases where rapidity cuts are larger than logarithms involving the veto scale, the considerations discussed here become relevant for phenomenology.

This work opens several possibilities for future investigations. Firstly, from a theoretical viewpoint, it is interesting to understand the perturbative structure of this observable at higher logarithmic accuracy in light of the several features discovered here.
It is also paramount to study such azimuthal vetoes in the context of the newly developed showers with higher logarithmic accuracy.
Finally, it will be important to study the potential impact of this class of resummations in phenomenological applications at the LHC, where accurate theoretical predictions are crucial to fully exploit the precise data.

\paragraph*{Acknowledgements.---}
We are grateful to Gavin Salam for helpful discussions and collaboration in early stages of this work, and in particular for pointing out the direct phenomenological implications of azimuthal vetoes, via the missing-$E_T$ isolation cuts used at the LHC. We further thank him for his comments and suggestions on the manuscript.
We thank the U.K.'s Science and Technologies Facilities Council for funding the project via grants ST/T001038/1 and ST/00077X/1 (MD), ST/X00077X/1 (SN), and via a Ph.D. studentship (AF).
%
The work of SN was also supported by the Royal Society through Grant No. URF/R1/201500.
The work of PM is funded by the European Union (ERC, grant agreement No. 101044599). Views and opinions expressed are however those of the authors only and do not necessarily reflect those of the European Union or the European Research Council Executive Agency. Neither the European Union nor the granting authority can be held responsible for them.

\paragraph*{Note Added.---}
While this Letter was being submitted, a new preprint~\cite{Banfi:2025mra} presented a calculation of CVLs for the one-jettiness variable. The NNDL order of CVLs observed in that article appears to be in line with our findings of Eq.~\eqref{eq:sigmaCVL}. However, more studies are necessary to understand the connection between the two results.

\bibliographystyle{apsrev4-2}
\bibliography{refs}

\newpage

\onecolumngrid
\newpage
\appendix


\setcounter{equation}{14}

\makeatletter
\renewcommand\@biblabel[1]{[#1S]}
\makeatother

\setcounter{secnumdepth}{3}

\section*{Supplemental material}
In this Appendix, we report some technical details needed to reproduce the results of the Letter.

\subsection{Non-global structure at NDL}
\label{app:NGLs-NDL}
At NDL accuracy it is necessary to consider the secondary branching of soft-collinear primary emissions, as well as the one-loop virtual corrections. For rIRC safe global observables the overall correction at NDL can be accounted for by use of the running coupling in the CMW scheme~\cite{Catani:1990rr,Banfi:2018mcq,Catani:2019rvy}. This treatment essentially combines the effect of virtual corrections with secondary splittings  \textit{inclusively}, i.e without imposing any additional constraints on the secondary partons 
compared to their parent. For non-global observables this treatment misses the possibility of soft wide-angle secondary radiation entering the veto region $\Omega$, with the parent outside. At NDL it is sufficient to consider the contribution of such configurations only at $\mathcal{O}(\bar{\alpha}^2)$, a calculation we perform explicitly below.

Since we are interested only in soft radiation, we can take $\hat{n}$ to align with the direction of the hard quark in each hemisphere, denoted $p_1$ and $p_2$ for definiteness. We consider the emission of two gluons $k_1$ and $k_2$ with $k_{t,1}\gg k_{t,2}$. The NDL contribution to the non-global correction is given by~\cite{Dasgupta:2001sh}
\begin{equation}\label{eq:NG_nlo}
    \Sigma^{\scriptscriptstyle{\rm (NDL)}}_{\scriptscriptstyle\text{NG}}=-\int\left[dk_1\right]\left[dk_2\right]\Theta\left(k_{t,1}-k_{t,2}\right)\Tilde{\mathcal{M}}^2(k_1,k_2)\left(\Theta\left(v_{\Omega}(k_1,k_2)-e^{-L}\right)-\Theta\left(v_{\Omega}(k_1+k_2)-e^{-L}\right)\right),
\end{equation}
with $\Tilde{\mathcal{M}}^2(k_1,k_2)\equiv\mathcal{M}^2(k_1,k_2)-\mathcal{M}^2(k_1)\mathcal{M}^2(k_2)$ the correlated part of the two emission matrix element and $v_\Omega~\equiv~E_{T,\Omega}/Q$. $v_\Omega(k_1,k_2)$ measures the combined transverse-energy flow into $\Omega$ of both daughters of the secondary branching, whilst $v_\Omega(k_1+k_2)$ measures only that of the parent prior to branching. For strong energy ordering $\Tilde{\mathcal{M}}^2$ reduces to
\begin{equation}
    \Tilde{\mathcal{M}}^2(k_1,k_2)= 8C_FC_AS
\end{equation}
with 
\begin{equation}
    S=\frac{(p_1p_2)}{(p_1k_1)(k_1p_2)}\left(\frac{(p_1k_1)}{(p_1k_2)(k_2k_1)}+\frac{(k_1p_2)}{(k_1k_2)(k_2p_2)}-\frac{(p_1p_2)}{(p_1k_2)(k_2p_2)}\right).
\end{equation}
Quark recoil effects can be neglected and so $p_1$ and $p_2$ can be approximated to be back to back. Expressed in terms of rapidity and azimuth with respect to $p_1$
\begin{equation}\label{NG integrand}
        [dk_1][dk_2]\Tilde{\mathcal{M}}^2(k_1,k_2)\approx8C_FC_A \bar{\alpha}^2 
        \left(\prod_{i=1,2}\frac{dv_i}{v_i}d\eta_i\frac{d\phi_i}{2\pi}\right)\frac{\cos(\phi_{12})}{\cosh(\eta)-\cos(\phi_{12})},
\end{equation}
with $v_i \equiv k_{ti}/Q$, $\phi_{12}\equiv \phi_1-\phi_2$ and $\eta\equiv\eta_1-\eta_2$. The strong energy ordering between $k_1$ and $k_2$ means that
\begin{equation}
    \Theta(v_{\Omega}(k_1+k_2)-e^{-L})\approx\Theta(v_{\Omega}(k_1)-e^{-L}).
\end{equation}
Eq.~\eqref{eq:NG_nlo} therefore only has weight when $k_2$ enters $\Omega$ whilst $k_1$ does not. It reads
\begin{equation}\label{wtaNG}
\Sigma^{\scriptscriptstyle{\rm (NDL)}}_{\scriptscriptstyle\text{NG}}=-2\cdot8C_FC_A\bar{\alpha}^2\int_{e^{-L}}^1 \frac{dv_1}{v_1}\int_{e^{-L}}^{v_1}\frac{dv_2}{v_2}
\int_0^{\log\frac{1}{v_1}}d\eta_1\int_0^{\log\frac{1}{v_2}}d\eta_2\int_{\notin\Omega}\frac{d\phi_1}{2\pi}\int_{\in\Omega}\frac{d\phi_2}{2\pi}\frac{\cos(\phi_{12})}{\cosh(\eta)-\cos(\phi_{12})},
\end{equation}
which we have expressed as twice the hemisphere contribution, accurate at NDL.
The integral can be simplified by using
\begin{equation}
    \begin{split}
        \int_0^{\log\frac{1}{v_1}}d\eta_1\int_0^{\log\frac{1}{v_2}}d\eta_2\frac{\cos(\phi_{12})}{\cosh(\eta)-\cos(\phi_{12})}
       &=\int_{-\infty}^{\infty}d\eta \int_0^{\log\frac{1}{v_1}}d\eta_1 \Theta\left(0<\eta_1-\eta<\log\frac{1}{v_2}\right)\frac{\cos(\phi_{12})}{\cosh(\eta)-\cos(\phi_{12})}\\
       &\approx \log\frac{1}{v_1} \int_{-\infty}^{\infty}d\eta \frac{\cos(\phi_{12})}{\cosh(\eta)-\cos(\phi_{12})}
    \end{split}
\end{equation}
where the final approximation reflects the fact that the integrand is dominated by $\eta\approx0$. Plugging this into Eq.~\eqref{wtaNG} and performing the integrals over $v_i$ reproduces Eqs.~\eqref{eq:NLO_ng_structure} and \eqref{eq:explicit_IngS}.

\subsection{Derivation of the DL resummation for the Thrust-axis case in $e^+e^-$}
\label{app:DL-thrust}
We work in the double–logarithmic (DL) limit, where emissions are soft–collinear and strongly ordered in both transverse momentum and rapidity. As discussed in the Letter, the observable has distinct dependence on the azimuthal width $\Delta\phi$ of the veto region $\Omega$ either side of $\Delta\phi = \pi$. This is due to the existence of a complementary region $\Omega_c$ azimuthally opposite to $\Omega$, which effectively doubles the azimuthal range over which the highest $k_t$ emission must be vetoed. For $\Delta\phi>\pi$ one needs to take care not to double count the overlap between $\Omega$ and $\Omega_c$. For simplicity we focus our derivation to the case that $\Delta\phi<\pi$, or equivalently $f=\Delta\phi/2\pi<1/2$. Writing
$\ell \equiv \ln(Q/k_t)$ and defining $v_\Omega\equiv E_{T,\Omega}/Q$, a natural \emph{naive} hypothesis is that the veto factorizes in the strongly ordered limit,
\begin{equation}\label{eq:veto_factorization}
\Theta\!\left(e^{-L}-v_{\Omega}\right)
\stackrel{\text{naive}}{\approx}\prod_i \Theta\!\left(e^{-L}-v_{\Omega}(k_i)\right),
\end{equation}
where we have denoted by $v_\Omega(k)$ the (normalized) transverse energy deposited in $\Omega$ by emission $k$. If this factorization held, DL resummation would reduce to a Sudakov suppression
\begin{equation}
\Sigma^{\scriptscriptstyle{\rm (DL)}}_{\scriptscriptstyle{\rm Thrust,naive}}=e^{-R(L)} ,
\end{equation}
with the (DL) radiator given by
\begin{equation}
R(L)=4 C_F  \int d\ell \,\bar{\alpha} \, \int_{-\ell}^{\ell} d\eta \int \frac{d\phi}{2\pi} \Theta\left(v_{\Omega}(k)-e^{-L}\right).
\end{equation}
However, once quark recoil effects are correctly included, Eq.~\eqref{eq:veto_factorization} fails even at DL accuracy for the present observable. At DL accuracy the full result factorizes into identical hemisphere contributions, so we work with $\Sigma^{\scriptscriptstyle{\rm (DL)}}_{\scriptscriptstyle{\rm Thrust,hemi}}$ and write
\begin{equation}
\Sigma^{\scriptscriptstyle{\rm (DL)}}_{\scriptscriptstyle{\rm Thrust}}=(\Sigma^{\scriptscriptstyle{\rm (DL)}}_{\scriptscriptstyle{\rm Thrust,hemi}})^{\,2}.
\end{equation}

\begin{figure}[h]
  \centering
  \subfigure[\,Naive DL veto]{%
    \includegraphics[width=0.35\textwidth]{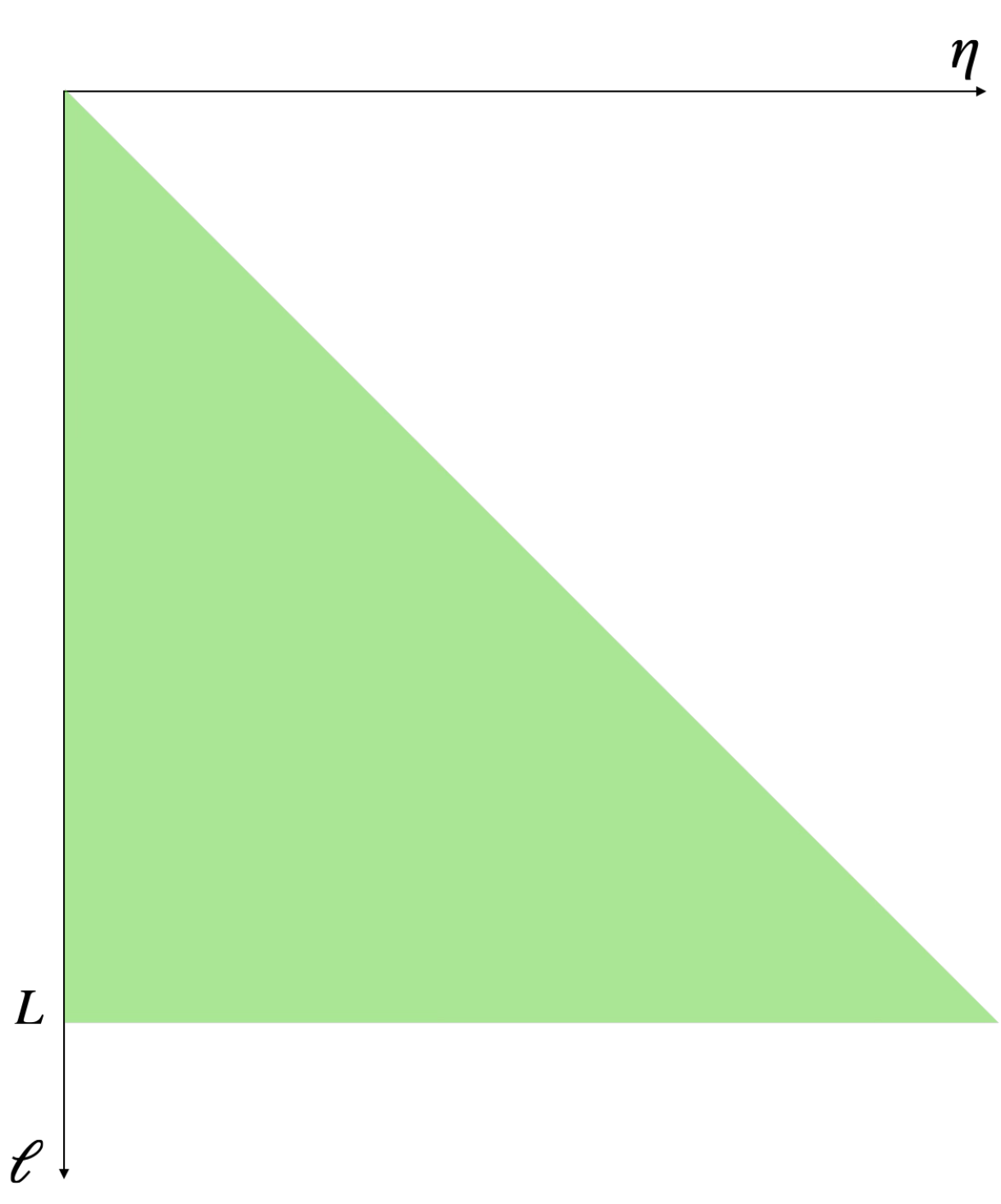}%
    \label{fig:thrust_naive}}
  \hspace{5em}%
  \subfigure[\,Correct veto for $k_1$ lowest $\ell$]{%
    \includegraphics[width=0.35\textwidth]{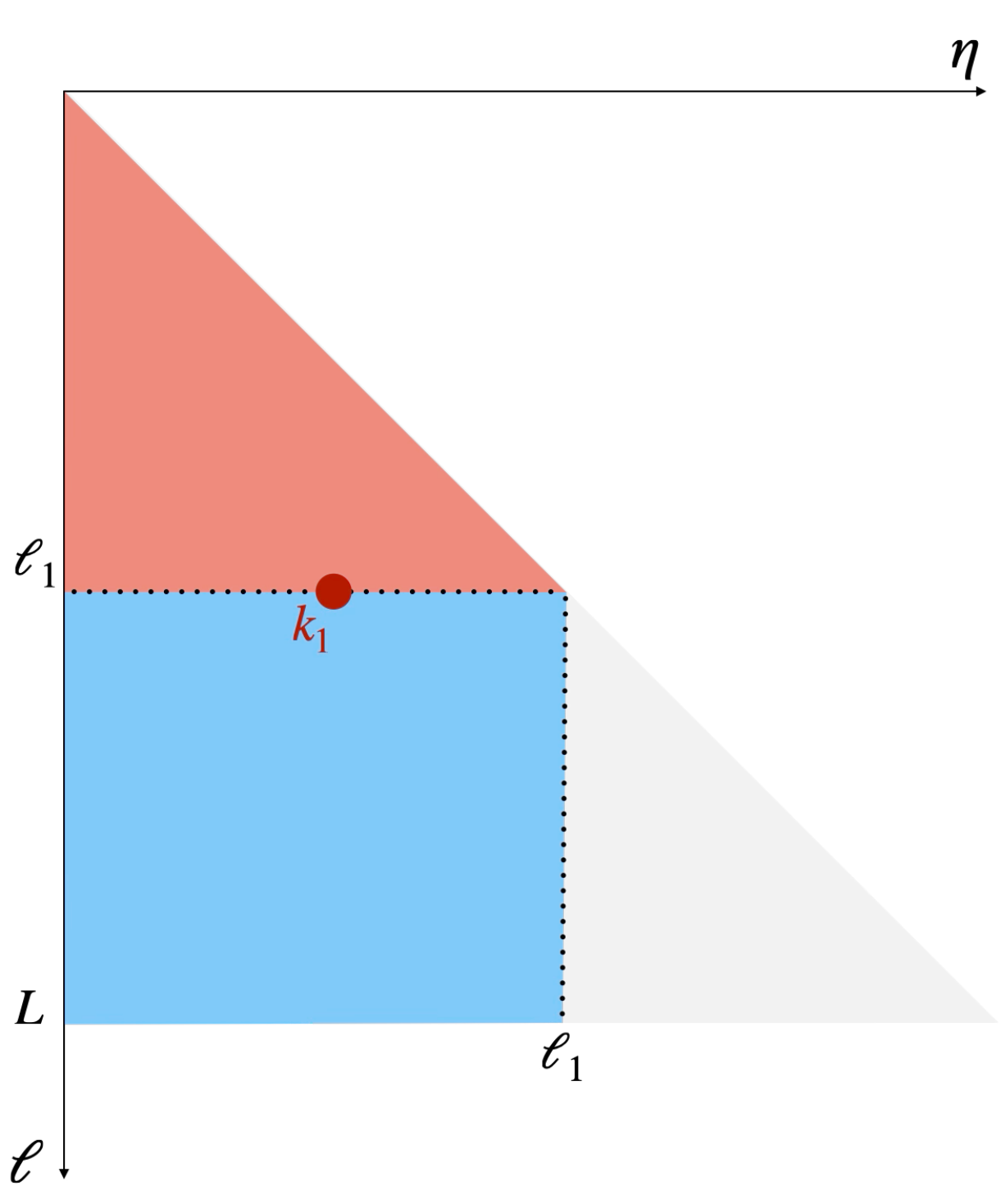}%
    \label{fig:thrust_hemi}}
  \caption{Lund plane for the DL veto in the thrust case. Green regions have azimuthal factor $2f$; blue regions $f$; red regions 1.}
  \label{fig:thrust_combo}
\end{figure}

We now examine the validity of Eq.~\eqref{eq:veto_factorization} for $\Omega$ an azimuthal strip of width $2\pi f$ and $\hat{n}$ aligned with the thrust axis. The one–gluon veto $\Theta\!\left(e^{-L}-v_{\Omega}(k)\right)$ forbids an emission with $\ell<L$ from entering $\Omega$, or the azimuthally opposite region $\Omega_c$ where it would recoil the quark into $\Omega$. For a single emission this acts only as an azimuthal constraint; the rapidity is unrestricted up to its kinematic bound.
The corresponding DL phase space is shown in Fig.~\ref{fig:thrust_naive} (green shading indicates an azimuthal factor $2f$).

Consider now a set of emissions that are strongly ordered in transverse momentum and rapidity, and label by $k_1$ the emission with the smallest $\ell$ (largest $k_t)$. Strong ordering ensures that only $k_1$ imparts a non–negligible recoil to the quark; $k_1$ thus obeys the same single–emission veto as described above, and displayed in Fig.~\ref{fig:thrust_naive}. The remaining emissions are \emph{not} subject to the naive product veto: (i) they are not prevented from entering $\Omega_c$, since they do not recoil the quark, and (ii) the quark–axis rapidity shift induced by $k_1$, $\eta_{nq}\simeq \ell_1$, removes any need for a veto on emissions with $\eta>\eta_{nq}$ (see Fig.~\ref{fig:tilt}). The resulting DL phase space is the blue region in Fig.~\ref{fig:thrust_hemi} (azimuthal factor $f$).

To resum at DL accuracy, we organize $\Sigma^{\scriptscriptstyle{\rm (DL)}}_{\scriptscriptstyle{\rm Thrust,hemi}}$ as the sum of a \emph{naive} Sudakov term—built from the single–emission radiator under the (incorrect) factorized veto—and a correction that starts at $\mathcal{O}(\alpha_s^2)$ and restores the recoil–induced phase space discussed above. Explicitly:

\begin{equation}\label{eq:thrust_DL_int}
\Sigma^{\scriptscriptstyle{\rm (DL)}}_{\scriptscriptstyle{\rm Thrust,hemi}}=e^{-R_{\text{hemi}}(0,L)}+4\bar{\alpha}C_F(1-2f)\int_0^Ld\ell_1 \int_0^{\ell_1} d\eta_1 \, e^{-R_{\text{global}}(0,\ell_1)}\left(e^{-R_{\text{true}}(\ell_1,L)}-e^{-R_{\text{hemi}}(\ell_1,L)}\right) ,
\end{equation}
with the $(1-2f)$ prefactor arising from the condition that emission $k_1$ is outside of both $\Omega$, and $\Omega_c$ where it would recoil the quark into $\Omega$. $R_{\scriptscriptstyle\text{hemi}}(L_1,L_2)$ is the single–emission radiator in one hemisphere,
\begin{equation}
R_{\scriptscriptstyle\text{hemi}}(L_1,L_2)
= 4 C_F\,(2f)\!\int_{L_1}^{L_2} \! d\ell\, \bar{\alpha} \!\int_{0}^{\ell} \! d\eta ,
\end{equation}
$e^{-R_{\text{global}}(0,\ell_1)}$ enforces that $k_1$ is the lowest–$\ell$ emission: a global veto on emissions with $\ell<\ell_1$ across all azimuth, corresponding to the red region in Fig.~\ref{fig:thrust_hemi}. The radiator is defined as
\begin{equation}
R_{\scriptscriptstyle\text{global}}(L_1,L_2)
= 4 C_F \!\int_{L_1}^{L_2} \! d\ell\, \bar{\alpha} \!\int_{0}^{\ell} \! d\eta\,.
\end{equation}
Finally $e^{-R_{\scriptscriptstyle\text{true}}(\ell_1,L)}$ implements the recoil–modified veto in the presence of $k_1$ (primary emissions restricted to $\eta<\ell_1$, cf. the blue region in Fig.~\ref{fig:thrust_hemi}), with
\begin{equation}
R_{\scriptscriptstyle\text{true}}(L_1,L_2)
= 4 C_F\, f \!\int_{L_1}^{L_2} \! d\ell\, \bar{\alpha} \!\int_{0}^{L_1} \! d\eta .
\end{equation}

\begin{figure}[h]
     \centering
     \includegraphics[width=0.5\linewidth]{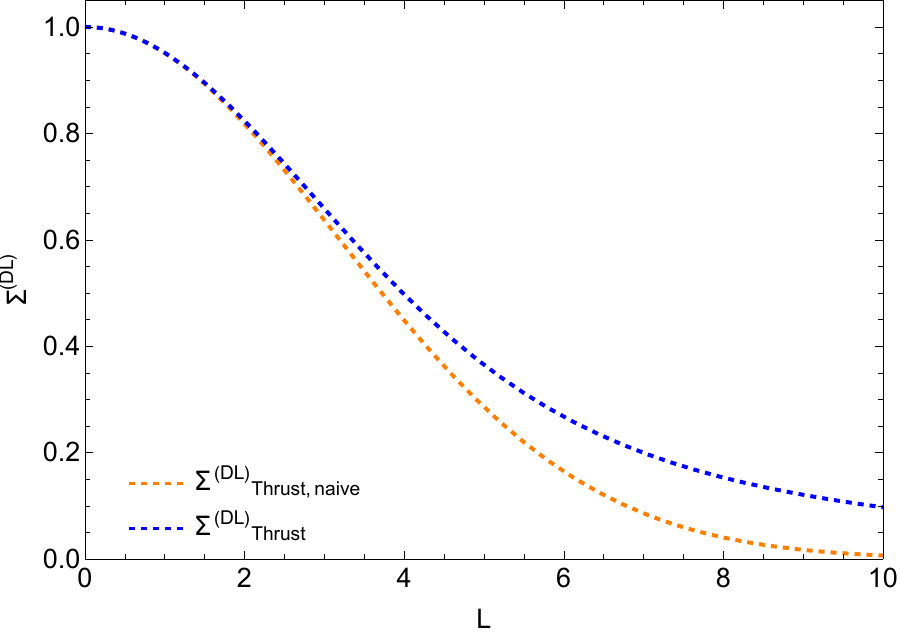}
     \caption{A comparison between the correct and naive DL resummations of $\Sigma^{\scriptscriptstyle{\rm (DL)}}_{\scriptscriptstyle{\rm Thrust}}$ for $f=1/4$ evaluated at $\bar{\alpha}=0.118/2\pi$.}
     \label{fig:thrust_resummation}
\end{figure}

Performing these integrals, and squaring $\Sigma^{\scriptscriptstyle{\rm (DL)}}_{\scriptscriptstyle{\rm Thrust,hemi}}$, we obtain

\begin{equation}
\Sigma^{\scriptscriptstyle{\rm (DL)}}_{\scriptscriptstyle{\rm Thrust}}=\frac{\Bigl(
  1 - 2f
  + \exp\!\left(\kappa \,f^{2}L^{2}\right)
    \sqrt{\kappa(1 - 2f)^2}\; f L \sqrt{\pi}\,
    \Bigl[
      \operatorname{erf}\!\Bigl(-\sqrt{\kappa}\,(1-f)\,L\Bigr)
      + \operatorname{erf}\!\Bigl(\sqrt{\kappa}\,f\,L\Bigr)
    \Bigr]
\Bigr)^{2}}{(1 - 2f)^{2}}\,,
\end{equation}
where we have defined $\kappa\equiv2\bar{\alpha}C_F/(1-2f)$. Note that this is a strictly DL result, and as such should be understood in the fixed coupling approximation. We plot the DL prediction $\Sigma^{\scriptscriptstyle{\rm (DL)}}_{\scriptscriptstyle{\rm Thrust}}$ alongside the naive exponentiation $\Sigma^{\scriptscriptstyle{\rm (DL)}}_{\scriptscriptstyle{\rm Thrust, naive}}$ in Fig.~\ref{fig:thrust_resummation}.

\subsection{Derivation of the NDL resummation for the WTA-axis case in $e^+e^-$}
\label{app:DL-WTA}
\begin{figure}[h]
  \centering
  \subfigure[\,Naive NDL veto]{%
    \includegraphics[width=0.35\linewidth]{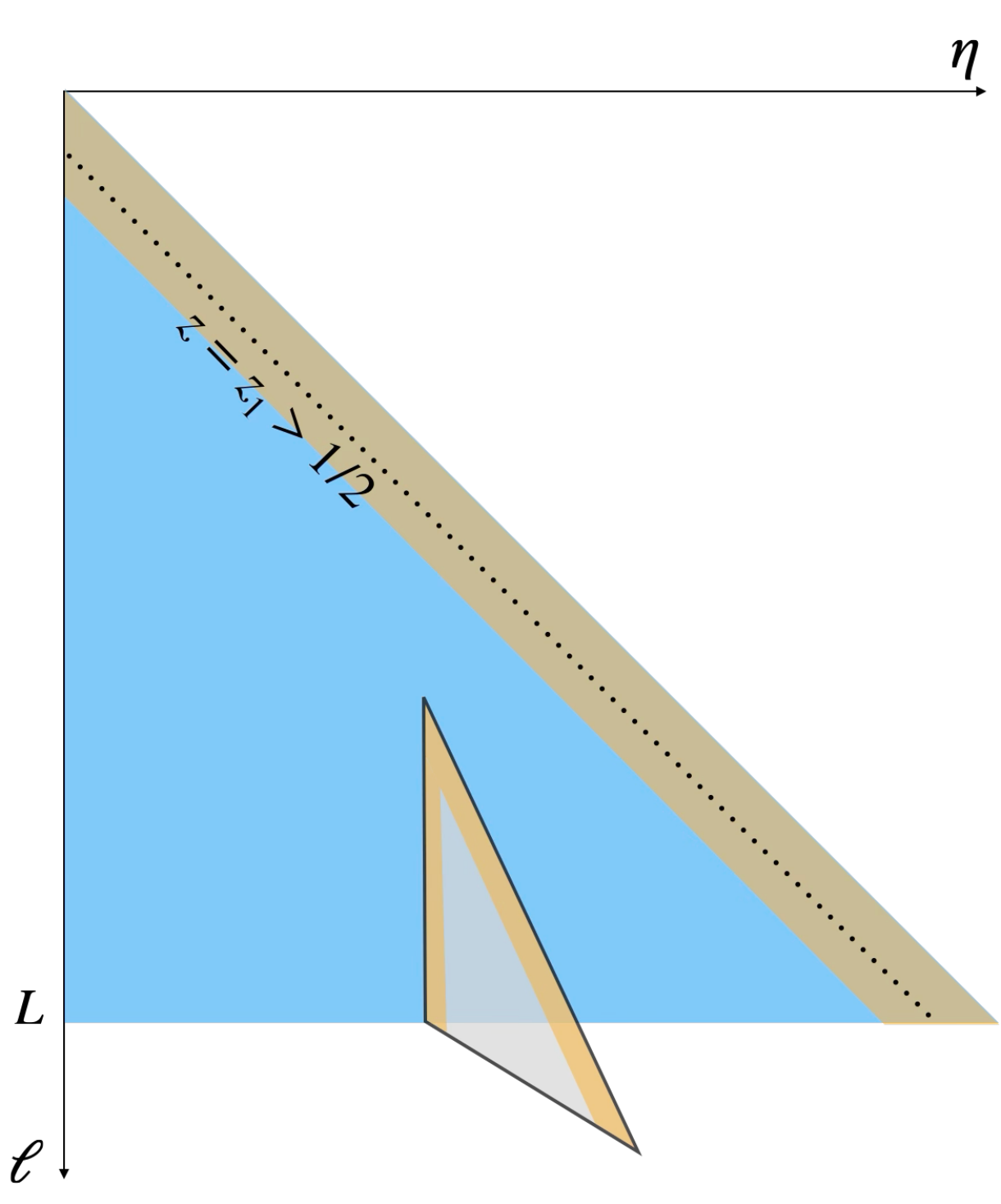}%
    \label{fig:wta_naive}}
  \hspace{5em}
  \subfigure[\,Correct NDL veto when $k_1$ (highest $z$) triggers axis swap]{%
    \includegraphics[width=0.35\linewidth]{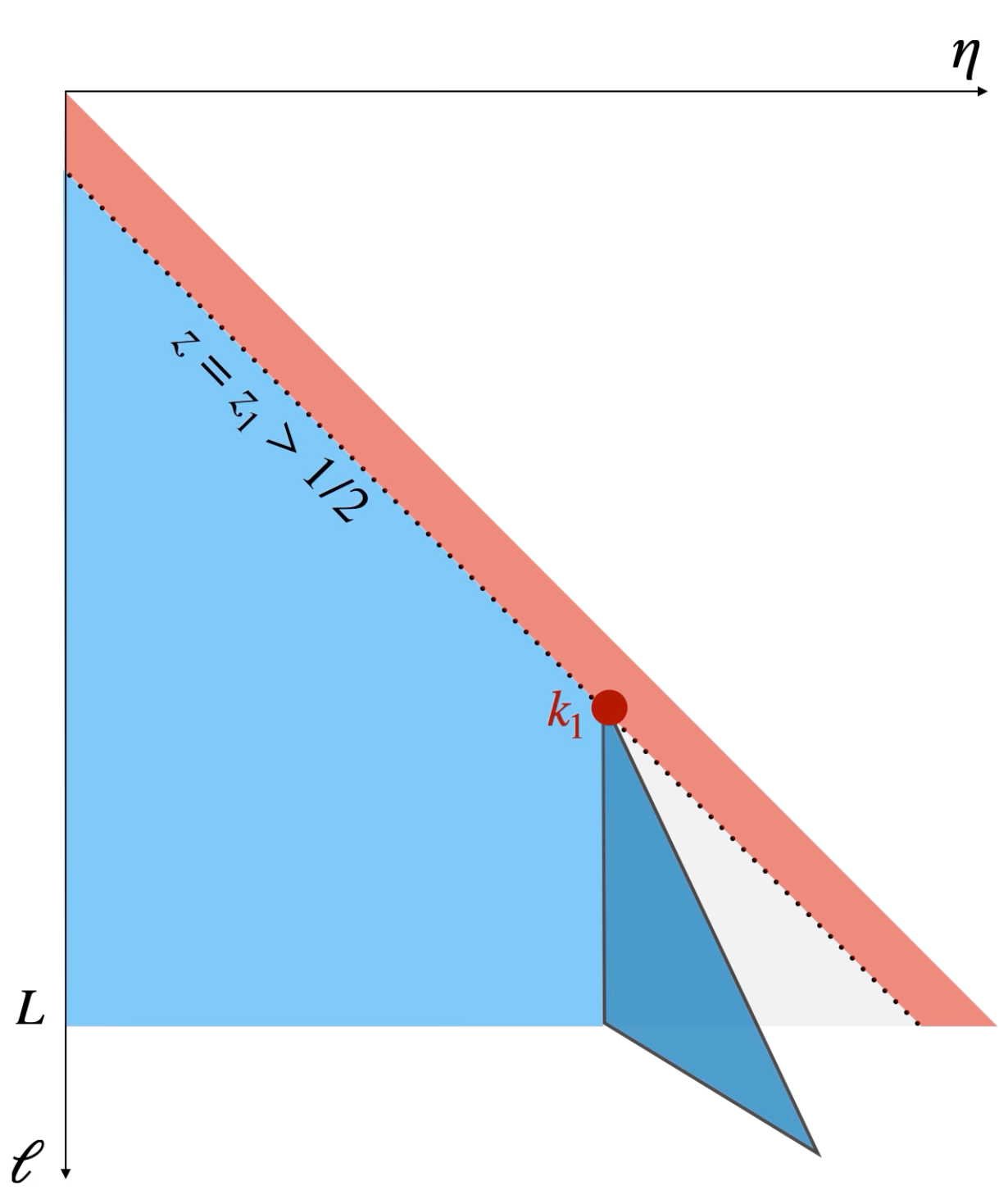}%
    \label{fig:wta_hemi}}
  \caption{Lund plane for the NDL veto in the WTA case. Blue regions have azimuthal factor $f$; red regions $1$. Gold shading indicates NDL corrections, without a specified azimuthal factor.}
  \label{fig:wta_combo}
\end{figure}

We now resum the WTA variant of $\Sigma$ to next-to-double-logarithmic (NDL) accuracy. As in the thrust case, the full result factorizes into identical hemisphere contributions, so we work with $\Sigma_{\scriptscriptstyle{\rm WTA, hemi}}$ and ultimately write~$\Sigma_{\scriptscriptstyle{\rm WTA}}~=~\Sigma_{\scriptscriptstyle{\rm WTA, hemi}}^{2}$. NDL accuracy requires control of kinematics in which \emph{at most one} parton is not strongly ordered in both $\ell$ and $\eta$. The corresponding single–emission radiator must therefore include (i) the hard–collinear refinement with the full $q\rightarrow qg$ splitting function~$P_{gq}=(1+(1-z^2))/z$, and the possibility of an \emph{axis swap}—for WTA this occurs for $z>1/2$ (equivalently $\eta>\ell-\ln 2$), where the axis aligns with the hard–collinear gluon—and (ii) NLO effects from secondary branchings through the running coupling together with non-global logarithms. The DL phase space is shown in Fig.~\ref{fig:wta_naive} as the blue ($f$–weighted) region of the primary Lund plane. The NDL ingredients are indicated in gold: the hard–collinear wedge of the primary Lund plane (full $P_{gq}$ and axis swap), and on the secondary Lund leaf, a gold hard–collinear region (running–coupling correction) together with a gold soft wide–angle region (non–global logarithms). Note that even if an emission has $z>1/2$ and so aligns with the WTA axis, it is still vetoed within an azimuthal region of size $2\pi f$ to ensure that the quark remains outside $\Omega$. If only these ingredients were relevant, one would obtain the “naive” NDL form

\begin{equation}\label{eq:NDL naive}
\Sigma^{\scriptscriptstyle{\rm (NDL)}}_{\scriptscriptstyle{\rm WTA,hemi}}=\Sigma^{\scriptscriptstyle{\rm (NDL)}}_{\scriptscriptstyle{\rm WTA,naive}}=\mathcal{S}\,e^{-R_{\text{hemi}}(0,L)},
\end{equation}
with the non-global factor $\mathcal{S}=1+S_2\,\bar{\alpha}^{\,2}L^{3}$ (see Eqs.~\eqref{eq:NLO_ng_structure} and \eqref{eq:explicit_IngS}) and the single emission radiator
\begin{equation}
R_{\scriptscriptstyle\text{hemi}}(L_1,L_2)
=2C_F\,f\!\int_{L_1}^{L_2}\! d\ell\;\frac{\alpha(Qe^{-\ell})}{2\pi}
\int_{0}^{\ell}\! d\eta\; z\,P_{gq}(z).
\end{equation}
Since both the running–coupling correction and the non–global contribution first enter at NDL, any interplay between these effects and the axis–swap (hard–collinear) sector is beyond our accuracy. Equivalently, the secondary Lund leaf in Fig.~\ref{fig:wta_naive} should be seeded only by emissions from the DL domain of the primary plane; it does not require additional seeding from the hard-collinear (including the axis-swap) wedge.

The naive expression in Eq.~\eqref{eq:NDL naive} assumes that the veto factorizes even in the presence of an axis-swapping emission. In other words, it presumes that strongly ordered soft–collinear emissions are insensitive to an axis swap — an assumption we now show to be incorrect. The highest energy emission is labeled $k_1$ with $z_1>1/2$ (equivalently $\eta_1>\ell_1-\ln 2$). The correct veto on the \emph{other} emissions is modified by the axis choice: the quark–axis rapidity tilt $\eta_{nq}=\eta_1$ removes any veto on primary emissions with $\eta>\eta_1$. The resulting primary phase space restriction is the blue, $f$–weighted region in Fig.~\ref{fig:wta_hemi}. Because the WTA axis is now defined by the hard gluon $k_1$, emissions that are doubly strongly ordered and emitted off $k_1$ (soft–collinear to $k_1$) must also be vetoed from entering $\Omega$. The excluded region is the secondary “leaf’’ attached to $k_1$ on the Lund plane in Fig.~\ref{fig:wta_hemi} (again with azimuthal weight $f$).

\begin{figure}[h]
     \centering
     \includegraphics[width=0.5\linewidth]{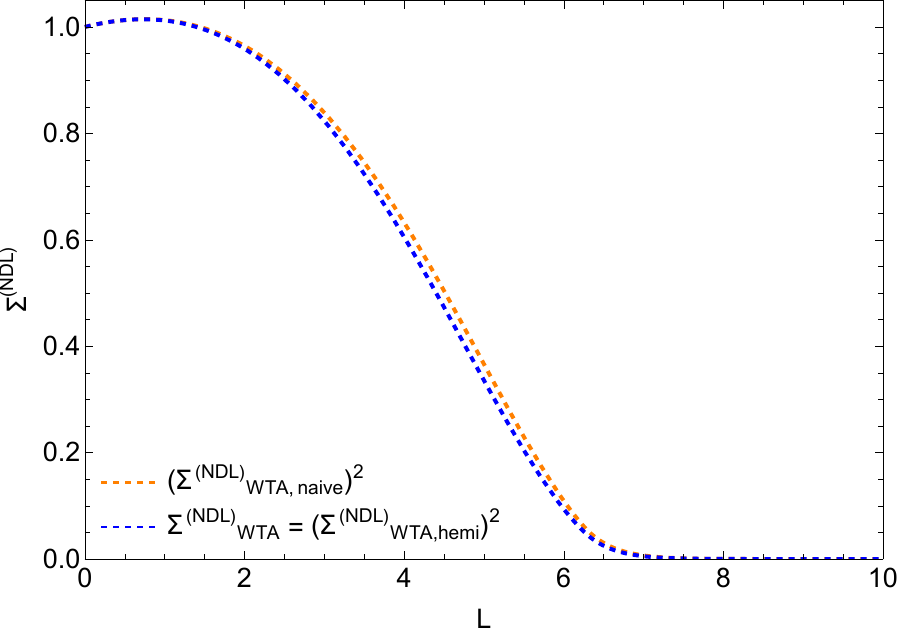}
     \caption{A comparison between the correct and naive NDL resummations of $\Sigma^{\scriptscriptstyle{\rm (NDL)}}_{\scriptscriptstyle{\rm WTA}}$ for $f=1/4$ evaluated at $Q=M_Z$.}
     \label{fig:wta_resummation}
\end{figure}

Collecting these ingredients, we organize the hemisphere result as
\begin{equation}
\Sigma^{\scriptscriptstyle{\rm (NDL)}}_{\scriptscriptstyle{\rm WTA,hemi}}=\Sigma^{\scriptscriptstyle{\rm (NDL)}}_{\scriptscriptstyle{\rm WTA,naive}}+\Sigma^{\scriptscriptstyle{\rm (NDL)}}_{\scriptscriptstyle{\rm WTA,correc}},
\end{equation}
with the correction that restores the non–factorising veto in axis–swap configurations,
\begin{equation}\label{eq:wta_hemi_correction}
\Sigma^{\scriptscriptstyle{\rm (NDL)}}_{\scriptscriptstyle{\rm WTA,correc}}
=2C_F(1-f)\!\int_{0}^{L}\! d\ell_1\,\bar{\alpha}
\int_{\ell_1-\ln 2}^{\ell_1}\! d\eta_1\; z_1 P_{gq}(z_1)\!
\left[
e^{-R_{\text{true},q}(\ell_1,0,L)}\,e^{-R_{\text{true},g}(\ell_1,\ell_1,L)}
- e^{-R_{\text{hemi}}(0,L)}
\right],
\end{equation}
where after changing coordinates~$z P_{gq}(z)=1+\left(1-e^{-(l-\eta)}\right)^2$. The $(1-f)$ prefactor reflects the azimuthal condition on $k_1$ to keep the quark out of $\Omega$. In contrast to Eq.~\eqref{eq:thrust_DL_int}, there is no $e^{-R_{\text{global}}}$ factor: enforcing $k_1$ as the highest energy emission would amount to a single–log veto (the red region in Fig.~\ref{fig:wta_hemi}) and is subleading at NDL. The first product of exponentials implements the correct veto in the presence of an axis swap (blue regions on the primary and secondary Lund planes in Fig.~\ref{fig:wta_hemi}), while the last term subtracts the naive factorized veto. Explicitly,
\begin{equation}
R_{{\scriptscriptstyle \text{true}},q}(\tilde{\eta},L_1,L_2)
=4C_F\,f \!\int_{L_1}^{L_2}\! d\ell\,\bar{\alpha}\!\int_{0}^{\min(\ell,\tilde{\eta})}\! d\eta,
\end{equation}
and
\begin{equation}
R_{{\scriptscriptstyle\text{true}},g}(\tilde{\eta},L_1,L_2)
=4C_A\,f \!\int_{L_1}^{L_2}\! d\ell\,\bar{\alpha}\!\int_{\tilde{\eta}}^{\ell}\! d\eta.
\end{equation}
We have used $\ell_1$ rather than $\eta_1$ as argument in Eq.~\eqref{eq:wta_hemi_correction}, which is sufficient at NDL accuracy. Since $\Sigma^{\scriptscriptstyle{\rm (NDL)}}_{\scriptscriptstyle{\rm WTA,correc}}$ is pure NDL, the full result to this accuracy is
\begin{equation}\label{eq:NDL_structure_wta}
\Sigma^{\scriptscriptstyle{\rm (NDL)}}_{\scriptscriptstyle{\rm WTA}}=\left(\Sigma^{\scriptscriptstyle{\rm (NDL)}}_{\scriptscriptstyle{\rm WTA,hemi}}\right)^{2}
\simeq \left(\Sigma^{\scriptscriptstyle{\rm (NDL)}}_{\scriptscriptstyle{\rm WTA,naive}}\right)^{2}
+2\,\Sigma^{\scriptscriptstyle{\rm (NDL)}}_{\scriptscriptstyle{\rm WTA,naive}}\,\Sigma^{\scriptscriptstyle{\rm (NDL)}}_{\scriptscriptstyle{\rm WTA,correc}}.
\end{equation}
Plugging the above radiators into Eq.~\eqref{eq:NDL_structure_wta} and performing the integrals yields the result quoted in Eq.~\eqref{Eq:pp resummation result}. The functions $\Sigma^{\scriptscriptstyle{\rm (NDL)}}_{\scriptscriptstyle{\rm WTA}}=\left(\Sigma^{\scriptscriptstyle{\rm (NDL)}}_{\scriptscriptstyle{\rm WTA,hemi}}\right)^{2}$ and $\left(\Sigma^{\scriptscriptstyle{\rm (NDL)}}_{\scriptscriptstyle{\rm WTA,naive}}\right)^2$ are shown in Fig.~\ref{fig:wta_resummation}.

\subsection{Fixed order computation of coherence-violating logarithms in Drell-Yan production}
\label{app:CVL}
In this appendix, we address the computation of the leading coherence-violating logarithms (CVLs) in the hadron-collider case. We work with the Drell-Yan process, as done in the Letter.
The calculation of the leading contribution can be carried out following the computation of its counterpart in the rapidity-gap-between-jets problem in Refs.~\cite{Forshaw:2008cq,Becher:2021zkk,Becher:2023mtx}. In both cases, the leading term of the coherence-violating tower of logarithmic corrections appears at ${\cal O}(\as^5)$. In the rapidity gap case, the corresponding configuration is given by two soft-collinear emissions along the beam direction, the exchange of two Glauber gluons, and the emission of a soft gluon with a large angle that enters the rapidity gap and gets vetoed. This configuration results in an $\as^5 L^7$ term that, in the rapidity gap case, is super leading given the single-logarithmic nature of the standard non-global series. A very similar configuration gives rise to a coherence-violating correction in the azimuthal gap case. However, in this case, the vetoed gluon can be both soft \textit{and} collinear, leading to a further logarithmic enhancement of this correction, hence resulting in an $\as^5 L^8$ contribution. An NNDL coherence-violating correction has never been encountered before for non-global observables.

To demonstrate the presence of such a term we perform an explicit $\alpha_s^5 $ calculation for the leading CVL in this case. We shall work with transverse momenta and rapidities defined with respect to the beam direction. We label the incoming partons as $k_1$ and $k_2$, and we associate the positive rapidity direction to that of parton $k_1$. The two out-of-gap soft-collinear emissions are labeled as $k_3$ and $k_4$, and the in-gap soft-collinear emission as $k$. Their transverse momenta are ordered as $k_{t,3}\gg k_{t,4}\gg k_{t}\gg E_T$. The rapidities of the emissions are only kinematically constrained, such that $|\eta_{i}|<\ymax{i}=\ln({Q}/{k_{t,i}})$. We denote by $\hat{\sigma}_{\rm CVL}$ the leading partonic cross section for the CVL correction differential in the kinematics of the Drell-Yan pair.
This is given by~\cite{Forshaw:2008cq}
\begin{align}
\label{eq:sigma-2-simp-b}
    \hat{\sigma}_{\scriptscriptstyle{\rm CVL}} =&\; \left(-\asfact\right)^5 \left(1-f\right)^2 
    \int_{E_T}^Q \frac{dk_{t,3}}{k_{t,3}}
    \int_{-\ymax{3}}^{\ymax{3}} d\eta_3
    \int_{E_T}^{k_{t,3}}\frac{dk_{t,4}}{k_{t,4}}
    \int_{-\ymax{4}}^{\ymax{4}} d\eta_4\int_{E_T}^{k_{t,4}}\frac{dk_t}{k_t}\nonumber\\
    &\qquad
  \bigg[\frac{1}{3}\ln^2\left(\frac{k_{t,4}}{E_T} \right)\langle m_0|  \D_{a_3,(2)}^{\mu_3\,\dagger}\D_{a_4,(3)}^{\mu_4\,\dagger}  \com{\G_{(4)}^\mathrm{I}}{\com{\G_{(4)}^\mathrm{I}}{\G_{(4)}^\mathrm{R}}}   \D_{a_4,(3)}^{\mu_4}\D_{a_3,(2)}^{\mu_3}
    |m_0\rangle\bigg]+{\cal O}(\alpha_s^6).
\end{align}
In Eq.~\eqref{eq:sigma-2-simp-b}, the prefactor $(1-f)^2$ arises from requiring that the emissions $k_3$ and $k_4$ are out of the azimuthal gap $\Omega$, while the factor $\ln^2\left(k_{t,4}/E_T\right)$ is the result of the integral over the transverse momentum of the two Glauber gluons between $E_T$ and $k_{t,4}$. We use the color-space formalism~\cite{Catani:1996vz} and denote the Born amplitude by $|m_0\rangle$, such that $\langle m_0|m_0\rangle = |m_0|^2$.
Following Ref.~\cite{Forshaw:2008cq} we have defined the real part of the anomalous dimension matrix $\G_{(4)}^\mathrm{R}$ as
\begin{equation} \label{eq:real-Gamma}
    \G^\mathrm{R}_{(4)}=-\frac{1}{2}\sum_{i<j}^4\T_{i,(4)} \cdot \T_{j,(4)} \int_{k\in\Omega}\frac{d\eta\, d\phi}{2\pi}\omega_{ij},
\end{equation}
where 
\begin{align}
\label{eq:omega-ij}
    \omega_{ij}&=\omega_{ji}=\frac{1}{2}k_t^2 \frac{k_i \cdot k_j}{(k \cdot k_i)(k \cdot k_j)}\approx  \Theta(\eta_i-\eta)\,\Theta(\eta-\eta_j)+\Theta(\eta_j-\eta)\,\Theta(\eta-\eta_i),
\end{align}
the last approximation being sufficient to capture the leading-logarithmic contributions. The color matrix $\T_{i,(n)}$ corresponds to a gluon emission from leg $i$ in an $n$-parton system. It is straightforward to show that in our case
\begin{align}
    \G_{(4)}^\mathrm{R} &= f\bigg[C_F\, \ln\left(\frac{Q}{k_t}\right)+\Theta(\eta_3-\eta_4)(\eta_3-\eta_4)\left(\frac{C_A}{2}+\T_{1,(4)}\cdot \T_{3,(4)}\right)\nonumber\\&\qquad+\Theta(\eta_4-\eta_3)(\eta_4-\eta_3)\left(\frac{C_A}{2}+\T_{1,(4)}\cdot \T_{4,(4)}\right) \bigg].
\end{align}
The terms proportional to $C_{F,A}$ in the above equation vanish inside the commutator in Eq.~\eqref{eq:sigma-2-simp-b}, and so they can be dropped. One can now freely perform the $k_t$ integral to obtain another power of $\ln (k_{t,4}/E_T)$. The real emission matrices $\D_{a_\ell,(n)}^{\mu}$ are defined as
\begin{equation}
\label{eq:real-emission-matrix}
\D_{a_\ell,(n)}^{\mu} = \sum_{i=1}^{n}\T_{i,(n)}^{a_{\ell}}h_{i,\ell}^{\mu}, \qquad
h_{i,\ell}^\mu = \frac{1}{2}k_{t,\ell} \frac{k_i^\mu}{k_{i}\cdot k_\ell},
\end{equation}
and have the properties
\begin{align}
\label{eq:D2-color-simp}
     \D_{a_3,(2)}^{\mu_3\,\dagger}\X\,
    \D_{a_3,(2)}^{\mu_3} &=-C_F \X,\\[5pt]
    \D_{a_4,(3)}^{\mu_4\,\dagger}\X
    \D_{a_4,(3)}^{\mu_4} &=-\T_{1,(3)}^{a_4\,\dagger} \X \,\T_{1,(3)}^{a_4} \,\Theta(\eta_4-\eta_3)-\T_{2,(3)}^{a_4\,\dagger} \X \,\T_{2,(3)}^{a_4}\, \Theta(\eta_3-\eta_4)\,,
\end{align}
where it is understood that the above color matrices act on color singlet states (amplitudes).
Finally, the imaginary part of the anomalous dimension matrix arising from Glauber gluons is
\begin{equation}
    \label{eq:imaginary-Gamma}
    \G^\mathrm{I}_{(n)}=i \pi\, \T_{1,(n)}\cdot\T_{2,(n)}+\frac{i \pi}{4}\left(\T_{1,(n)}^2+\T_{2,(n)}^2-\sum_{i=3}^n \T_{i,(n)}^2\right)\,,
\end{equation}
where the terms in round brackets proportional to quadratic Casimir operators can be dropped since they have no physical effect.
The color algebra in \EQ\eqref{eq:sigma-2-simp-b} can be performed with the help of the \texttt{ColorMath} package~\cite{Sjodahl:2012nk}, and we obtain
\begin{align}
\label{eq:color-structure-simplification}
    \langle m_0|  \D_{a_3,(2)}^{\mu_3\,\dagger}\D_{a_4,(3)}^{\mu_4\,\dagger}  \com{\G_{(4)}^\mathrm{I}}{\com{\G_{(4)}^\mathrm{I}}{\G_{(4)}^\mathrm{R}}}   \D_{a_4,(3)}^{\mu_4}\D_{a_3,(2)}^{\mu_3}
    |m_0\rangle&=\pi^2C_F\frac{N_c^2}{8}f\, |m_0|^2 \big|\eta_4-\eta_3\big|.
\end{align}
After performing the remaining kinematic integrals, including the convolution with the parton densities, we reach the final result~\eqref{eq:sigmaCVL}
\begin{align}
\label{eq:final-sigma-2-result}
    \Sigma^{\scriptscriptstyle{\rm (CVL)}}_{\scriptscriptstyle{\rm DY}} = - \left(\frac{\as}{\pi}\right)^5 \pi^2 C_F N_c^2 f\,\left(1-f\right)^2\frac{L^8}{180},
\end{align}
where  $L=\ln(Q/E_T)$.
Note that the result has an extra factor of $L$ compared to the rapidity gap computation, which, as mentioned earlier, gives $\as^{5}L^{7}$~\cite{Becher:2021zkk,Becher:2023mtx}. This is directly linked to the presence of the rapidity difference $\big|\eta_4-\eta_3\big|$  in \EQ\eqref{eq:color-structure-simplification} which gives an extra log factor $L$ upon performing the rapidity integrals.

\subsection{An algorithm for the LL resummation in Drell-Yan production}
\label{app:NGL-LL}
At LL, the $E_{T,\Omega}$ distribution in the Drell-Yan process can be obtained by solving the LL evolution equation given in Refs.~\cite{Banfi:2002hw,Banfi:2021owj} complemented with the physical rapidity bound, in the dipole frame, $|\eta_i| \leq \ln Q/k_{t,i}$, where $k_{t,i}$ denotes the dipole transverse momentum of the $i$-th emission in the evolution.
A Markov-chain Monte Carlo algorithm to solve such an equation can be obtained by adapting the algorithm first presented in Ref.~\cite{Dasgupta:2001sh}, and formulated for the $k_t$-ordered evolution in Ref.~\cite{Banfi:2021owj}. We use the latter version as a starting point, and define the evolution time
\begin{equation}
\label{eq:time}
t_i = \int_{\frac{k_{t,i}}{Q}}^1\frac{d x}{x} N_c \frac{\alpha_s(x Q)}{\pi} =-
\frac{N_c}{2\pi\beta_0}\ln \left(1-2 \beta_0\lambda_i\right),\qquad \lambda_i=\alpha_s(Q)\ln\frac{Q}{k_{t,i}}.
\end{equation}
We also define the infrared cutoff scale $Q_0 > \Lambda$ with $\Lambda$ being the
singularity of Eq.~\eqref{eq:time}, namely
\begin{equation}
\label{eq:Q0def}
2\,\beta_0\alpha_s(Q)\ln\frac{Q}{\Lambda} = 1,\quad \beta_0= \frac{11 C_A-2 n_f }{12\pi}.
\end{equation}
In practice, this scale is extremely low and therefore the evolution is very rarely terminated because the scale $Q_0$ is reached before any emission is radiated into the azimuthal gap $\Omega$. The LL resummation for the quantity $\frac{1}{\sigma}\frac{d \sigma }{d t}$ is obtained by iterating Algorithm~\eqref{algo:LLevolution} $N_{\scriptscriptstyle\rm events}$ times until the target statistical precision is reached. The distribution $\Sigma$ is obtained by calculating the cumulative integral of the resulting histogram.

An important aspect to stress about Algorithm~\eqref{algo:LLevolution}
is that at every step, emissions are generated in the emitting dipole
rest frame.
We therefore need to apply a simple Lorentz transformation at every
evolution step to transform the generated emission into the event
frame.
In doing this, we only need to keep track of the direction of the
generated momenta, while the information regarding the \textit{dipole}
transverse momentum (i.e.~the normalization of the momenta) is encoded
in the evolution time $t$.
Therefore, we divide all momenta by their energy in the event frame,
and keep track of the normalization separately.
As in Refs.~\cite{Dasgupta:2001sh,Banfi:2021owj}, all Lorentz transformation have to be
performed with \textit{normalized} momenta.
This solves the problem of handling numerically Lorentz
transformations involving very soft momenta. Also, it allows us to
perform the evolution without any momentum conservation at any
evolution step, thus eliminating exactly all subleading power
(i.e.~non-logarithmic) corrections.

The value of the rapidity cut $\eta_{\scriptscriptstyle\rm M}$ in Algorithm~\eqref{algo:LLevolution} must be such that the resummed logarithm $L < \eta_{\scriptscriptstyle\rm M}$. When $L \sim \eta_{\scriptscriptstyle\rm M}$ cutoff effects start affecting the physical distribution of the observable and one needs to increase $\eta_{\scriptscriptstyle\rm M}$ accordingly. The physical upper bound in the emission rapidity implemented in Algorithm~\eqref{algo:LLevolution} implies that the algorithm, accurate at LL, generates also small subleading logarithmic corrections to $\Sigma$. In order to neglect these in the final result, we follow the method of Ref.~\cite{Dasgupta:2020fwr} and take numerically the limit $\alpha_s\to 0$. Specifically, Fig.~\ref{fig:gnole} shows the quantity
\begin{equation}
   \lambda \,g_{\scriptscriptstyle{\rm LL}}(\lambda) = \lim_{\alpha_s\to 0} \alpha_s \ln \Sigma.
\end{equation}

\begin{algorithm}[H]
\caption{Generation of an event for $\frac{1}{\sigma}\frac{d \sigma }{d t}$}
\label{algo:LLevolution}
\begin{algorithmic}
   \State Set $i=0$ and the evolution time $t_0=0$;
   \State Start with one initial dipole ${\cal D}_1$ made by the Born $q\bar{q}$ line. Assign to the dipole a collinear cutoff in the dipole rest frame so that the magnitude of the rapidity of an emission with respect to the dipole extremities is less than $\eta_{\scriptscriptstyle\rm M}$. This determines the allowed rapidity span $\Delta\eta_{1}$ of the dipole in the event frame;
   \While{{\rm true}}
     \State Compute $\Delta\eta_{\scriptscriptstyle{\rm tot}}=\sum_{\ell=1}^N\Delta\eta_{\ell}$, the sum of the available event-frame rapidity
      ranges within each of the $N$ dipoles in the event so far;
     \State Increase $i$ by 1;
     \State Choose the emitting dipole ${\cal D}_\ell$ with probability $\Delta\eta_\ell/\Delta\eta_{\scriptscriptstyle{\rm tot}}$;
     \State Generate a random number $r \in [0,1]$ \& increase $t$ by an amount $\Delta t = t_i-t_{i-1}$ generated by solving
     \begin{equation}
         e^{-\Delta\eta_{\scriptscriptstyle{\rm tot}}\Delta t} = r\,;
     \end{equation}
    \State Generate the \textit{dipole} transverse momentum $k_{t,i}$ of the next emission $k_i$ by solving Eq.~\eqref{eq:time} with the new
   $t_i$. Generate its azimuth $\phi_i$ and rapidity $\eta_i$ uniformly in the dipole frame with $\phi_i\in[0,2 \pi]$ and $|\eta_i|\leq \ln\frac{Q}{k_{t,i}}$, espectively;
     \If{$|\eta_i| > \eta_{\scriptscriptstyle\rm M}$} 
     \State Discard the \textit{emission} and generate a new one starting from scale $t_i$;
     \EndIf
     \If{$k_{t,i} < Q_0$} 
     \State Stop the evolution and generate a new event;
     \EndIf
     \State Split the dipole ${\cal D}_\ell$ into two adjacent dipoles, and assign each new dipole a collinear cutoff in its own rest frame so that the magnitude of the rapidity of an emission with respect to the dipole extremities is less than $\eta_{\scriptscriptstyle\rm M}$. Boost the rapidity bound into the event frame, to determine the rapidity span of each new dipole;
     %
     \If{$k_i\in\Omega$} 
     \State Add $w=1/N_{\scriptscriptstyle{\rm events}}$ to the $t_i$ bin in the histogram
       \& generate a new event;
     \EndIf
   \EndWhile
\end{algorithmic}
\end{algorithm}

\subsection{Fixed-order study of the logarithmic accuracy of $k_t$-ordered local dipole showers}
\label{app:PS}
Here, we describe the calculation of the $\Delta \phi$ observable in the context of dipole showers with dipole-local recoil. It is well-known~\cite{Dasgupta:2018nvj} that the logarithmic accuracy of such showers is broken at the NLL level (or equivalently for single-logarithmic terms $\alpha_s^n L^n$) in the large-$N_c$ limit. The breaking of shower accuracy at NLL is due to issues in the attribution of transverse recoil and first arises for two soft emissions with commensurate $k_t$ and disparate rapidity. 

As we show below, this class of dipole showers, when applied to the case of the $\Delta \phi$ observable, fail to produce the correct logarithmic structure starting at $\alpha_s^2 L^3$ level, i.e. a leading logarithmic term in standard  $\ln \Sigma$ counting and an NDL term in $\Sigma$.  

In order to demonstrate this, we follow the approach of Ref.~\cite{Dasgupta:2018nvj} and carry out an explicit fixed-order analysis at $\alpha_s^2$, with two soft emissions widely separated in rapidity.  We label the emissions as $k_1$ and $k_2$ and take emission $k_1$ to be produced first in the shower. We also assume, for simplicity, a dipole-$k_t$ ordered shower so that $k_{t,1}>k_{t,2}$.    

The usual impact of incorrect recoil attribution is to modify the transverse momentum of emission $k_1$ by an order one factor due to the commensurate $k_t$ of emission $k_2$.
This usually leads to a single-logarithmic effect as shown in Ref.~\cite{Dasgupta:2018nvj}. In the present case the non-global nature of the observable, alongside a sensitivity to large rapidities $\eta \sim \ln Q^{\scriptscriptstyle\rm (dip)}/k_t$ (with $Q^{\scriptscriptstyle\rm (dip)}$ being the dipole invariant mass), causes a further complication which enhances the effect. Specifically, we can have an emission $k_1$ in the veto region which then moves out of the veto region through recoil, or a situation where both emissions are outside the veto region and emission $k_1$ is moved in after recoil. Hence, one finds a situation where the contribution to the observable is fundamentally affected rather than just modified by a factor of order one. 

Let us examine the effect of the above mentioned contributions by studying the differential distribution $\frac{v_{\Omega}}{\sigma} \frac{d\sigma}{dv_{\Omega}}$ where $v_{\Omega}\equiv E_{T,\Omega}/Q$.  We consider first a contribution where $k_1$ is outside the veto region $\Omega$ and is pushed into it after recoil. The emission $k_2$ is taken to be outside $\Omega$. When $k_1$ absorbs the transverse recoil from $k_2$ we have the following relation for the transverse momentum vectors :  ${\bf{k}}_{t,1} = {\bf{k}}^{\scriptscriptstyle\text{shower}}_{t,1}+ {\bf{k}}_{t,2}$, where ${\bf{k}}^{\scriptscriptstyle\text{shower}}_{t,1}$ denotes the emission $k_1$ after recoil against $k_2$.
Following the calculations of Ref.~\cite{Dasgupta:2018nvj}, we then have the following extra spurious contribution due to the recoil effect in the shower:
\begin{multline}
\label{eq:recoil}
\delta \left(\frac{v_{\Omega}}{\sigma} \frac{d\sigma}{dv_{\Omega}} \right)^{k_1, \text{in}} = v_{\Omega} \left(\frac{2 C_F \alpha_s}{\pi} \right)^2\,\int \frac{dv_{1} }{v_1} \int^{v_{1}}\frac{dv_{2}}{v_{2}} \int_{\ln v_1}^{\ln \frac{1}{v_1}} d \eta_1 \int_{\frac{1}{2}(\eta_1+\ln v_1)}^{\frac{1}{2}(\eta_1+\ln \frac{1}{v_1})}  d\eta_2\\ \int \frac{d\phi_1}{2\pi} \frac{d\phi_2}{2\pi} \delta(v_{\Omega}-v^{\scriptscriptstyle\text{shower}}_1)\Theta_{k^{\scriptscriptstyle\text{shower}}_1 \in \Omega}  \Theta_{k_1 \notin \Omega} \Theta_{k_2 \notin \Omega}.
\end{multline}

In the above equation $v_i = k_{t,i}/Q$, and the integral over $\eta_2$, the rapidity of $k_2$, is over the range where its recoil in the shower is absorbed by $k_1$, as discussed in Ref.~\cite{Dasgupta:2018nvj}. The constraints $\Theta_{k_1 \notin \Omega} \Theta_{k_2 \notin \Omega}$ are to be understood as conditions on the azimuthal angles such that the emissions $k_1$ and $k_2$ are out of the azimuthal strip $\Omega$. The constraint $\Theta_{k^{\text{shower}}_1 \in \Omega}$ is instead on the recoiled emission to be in $\Omega$. It depends on $\phi_1, \phi_2$ and the ratio $x=v_2/v_1$. We also note that in the delta function constraint, we have the condition $v_{\Omega}=v_1^{\scriptscriptstyle\text{shower}}$. The quantity $v_1^{\scriptscriptstyle\text{shower}}$ is given by 
\begin{equation}
v_1^{\scriptscriptstyle\text{shower}} =\sqrt{v_1^2+v_2^2-2 v_1 v_2 \cos (\phi_2-\phi_1)} = v_1 \sqrt{1+x^2 -2 x \cos (\phi_2-\phi_1)}.
\end{equation}
Up to sub-leading logarithmic corrections one may drop the factor multiplying $v_1$ above and the delta function constraint may accordingly be treated as $\delta(v_{\Omega}-v_1)$.
The rapidity integrals and the integral over $v_1$ are then trivial and give, for the difference between the shower and correct distributions
\begin{equation}
\delta \left(\frac{v_{\Omega}}{\sigma} \frac{d\sigma}{dv_{\Omega}} \right)^{k_1, \, {\scriptscriptstyle\text{in}}} =  \left(\frac{2 C_F \alpha_s}{\pi} \right)^2\ 2\, L^2 \,  \mathcal{F}_\Omega^{\scriptscriptstyle\text{in}},
\end{equation}
where $L=-\ln v_{\Omega}$ and we defined 
\begin{equation}\label{eq:F_in}
 \mathcal{F}_\Omega^{\scriptscriptstyle\text{in}} =\int \frac{dx}{x}\frac{d\phi_1}{2\pi} \frac{d\phi_2}{2\pi} \Theta_{k^{\text{shower}}_1 \in \Omega}  \Theta_{k_1 \notin \Omega} \Theta_{k_2 \notin \Omega}.
\end{equation}

A similar result applies to the case when emission $k_1$  is inside $\Omega$ and $k^{\scriptscriptstyle\text{shower}}_1$ and $k_2$ are outside. Here, the shower result for the differential distribution will vanish while the true result does not. The difference between the shower and correct result is 
\begin{multline}
\delta \left(\frac{v_{\Omega}}{\sigma} \frac{d\sigma}{dv_{\Omega}} \right)^{k_1, \, {\scriptscriptstyle\text{out}}} = -v_{\Omega} \left(\frac{2 C_F \alpha_s}{\pi} \right)^2\,\int \frac{dv_{1} }{v_1} \int^{v_{1}}\frac{dv_{2}}{v_{2}} \int_{\ln v_1}^{\ln \frac{1}{v_1}} d \eta_1 \int_{\frac{1}{2}(\eta_1+\ln v_1)}^{\frac{1}{2}(\eta_1+\ln \frac{1}{v_1})}  d\eta_2\\ \int \frac{d\phi_1}{2\pi} \frac{d\phi_2}{2\pi} \delta(v_{\Omega}-v^{\scriptscriptstyle\text{shower}}_1)\,\Theta_{k^{\scriptscriptstyle\text{shower}}_1 \notin \Omega}  \Theta_{k_1 \in \Omega} \Theta_{k_2 \notin \Omega},
\end{multline}
giving 
\begin{equation}
\delta \left(\frac{v_{\Omega}}{\sigma} \frac{d\sigma}{dv_{\Omega}} \right)^{k_1, \, {\scriptscriptstyle\text{out}}} =  \left(\frac{2 C_F \alpha_s}{\pi} \right)^2\ 2\, L^2 \,  \mathcal{F}_\Omega^{\scriptscriptstyle\text{out}},
\end{equation}
where we defined 
\begin{equation}\label{eq:F_out}
 \mathcal{F}_\Omega^{\scriptscriptstyle\text{out}}=- \int \frac{dx}{x}\frac{d\phi_1}{2\pi} \frac{d\phi_2}{2\pi} \Theta_{k^{\scriptscriptstyle\text{shower}}_1 \notin \Omega}  \Theta_{k_1 \in \Omega} \Theta_{k_2 \notin \Omega}.
\end{equation}

\begin{figure}[h]
     \centering
     \includegraphics[width=0.5\linewidth]{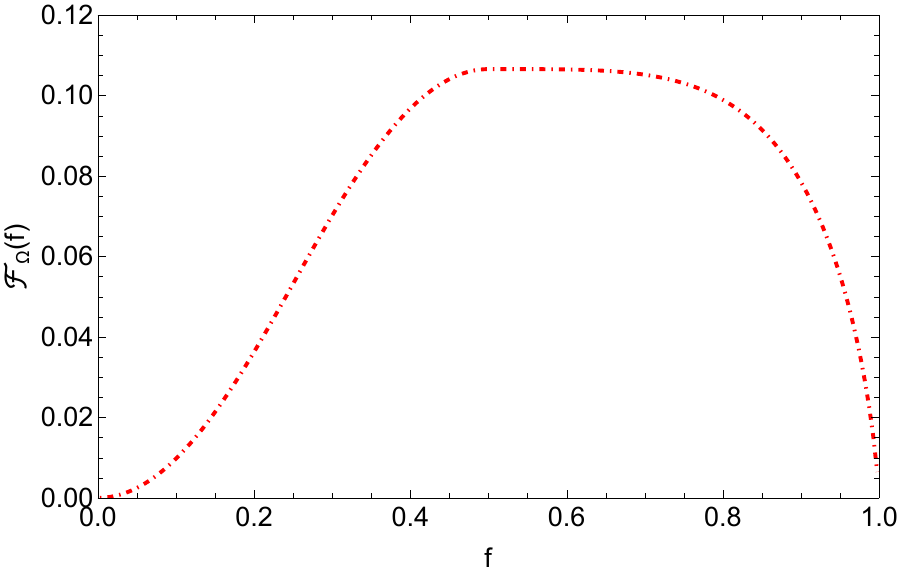}
     \caption{The quantity $\mathcal{F}_\Omega$ defined in Eqs.~\eqref{eq:recoil_error_result}, \eqref{eq:F_in} and \eqref{eq:F_out}.}
     \label{fig:F_Omega}
 \end{figure}

From the above, we see an $\alpha_s^2 L^2$ effect in $d\sigma/dL$ for the shower corresponding to an $\alpha_s^2 L^3$ term in $\Sigma$. The final overall result for the extra spurious shower contribution for the cumulative distribution $\Sigma$ is 
\begin{equation}\label{eq:recoil_error_result}
\delta \Sigma = \left(\frac{2 C_F \alpha_s}{\pi} \right)^2 \frac{2}{3} L^3 \,  \mathcal{F}_\Omega, \, \, \,  
\end{equation}
where $\mathcal{F}_\Omega =  \mathcal{F}_\Omega^{\scriptscriptstyle\text{out}}+ \mathcal{F}_\Omega^{\scriptscriptstyle\text{in}}$.
The quantity $\mathcal{F}_\Omega$ is evaluated straightforwardly with numerical integration. In order to carry this out, the integrals over $\phi_1$ and $\phi_2$ are performed over the relevant azimuthal ranges defined by a strip $\Omega$ extending between $-\pi f$ and $+\pi f$. The azimuthal constraint on $k_1^{\scriptscriptstyle\text{shower}}$ is a function of $\phi_1,\phi_2$ and $x$. Specifically, in a notation where we have ${\bf{k}}_t  = \left(k_{t,x},k_{t,y} \right ) = k_t (\cos \phi,\sin \phi)$, we can write 
\begin{equation}
\phi_1^{\scriptscriptstyle\text{shower}} =  \text{tan}^{-1} \left(\frac{k^{\scriptscriptstyle\text{shower}}_{t,y}}{k_{t,x}^{\scriptscriptstyle\text{shower}}}\right)  = \text{tan}^{-1} \frac{\sin \phi_1- x \sin\phi_2}{\cos \phi_1- x \cos \phi_2 },
\end{equation}
where we used ${\bf{k}}^{\scriptscriptstyle\text{shower}}_t = {\bf{k}}_{t1} - {\bf{k}}_{t2}$ and $x=k_{t,2}/k_{t,1}$.  The constraint for the shower emission $k^{\text{shower}}_1$ to be in the strip would involve the condition $\pi f>\phi_1^{\scriptscriptstyle\text{shower}} >- \pi f$, otherwise the emission is outside the strip. The quantity $\mathcal{F}_\Omega$ is shown in Fig.~\ref{fig:F_Omega}, showing that $k_t$-ordered dipole showers with a dipole-local recoil would produce an incorrect $\alpha_s^2 L^3$ term.

\end{document}